\documentclass[conference]{IEEEtran}
\IEEEoverridecommandlockouts
\usepackage{cite}
\usepackage{amsmath,amssymb,amsfonts}
\usepackage{graphicx}
\usepackage{textcomp}
\usepackage{xcolor}

\usepackage[most]{tcolorbox}
\tcbuselibrary{listings,theorems}
\tcbset{fonttitle=\bfseries}

\usepackage{url}
\usepackage{booktabs}
\usepackage{algpseudocode}
\usepackage[switch]{lineno}
\usepackage{verbatim} 
\usepackage{placeins}
\usepackage{pifont}
\usepackage{tcolorbox}
\usepackage{longtable}
\usepackage{graphicx}
\usepackage{subcaption}
\usepackage{booktabs}  
\usepackage{adjustbox} 
\usepackage{enumitem}
\usepackage{multirow}
\usepackage{soul}

\newcommand{\systemname}{ContextBuddy}

\def\BibTeX{{\rm B\kern-.05em{\sc i\kern-.025em b}\kern-.08em
    T\kern-.1667em\lower.7ex\hbox{E}\kern-.125emX}}
\begin{document}

\title{\systemname: AI-Enhanced Contextual Insights for Security Alert Investigation (Applied to Intrusion Detection)}

\author{
\IEEEauthorblockN{Ronal Singh}
\IEEEauthorblockA{\textit{Data61, CSIRO} \\
Melbourne, Australia}
\and
\IEEEauthorblockN{Mohan Baruwal Chhetri}
\IEEEauthorblockA{\textit{Data61, CSIRO} \\
Melbourne, Australia}
\and
\IEEEauthorblockN{Surya Nepal}
\IEEEauthorblockA{\textit{Data61, CSIRO} \\
Sydney, Australia}
\and
\IEEEauthorblockN{Cecile Paris}
\IEEEauthorblockA{\textit{Data61, CSIRO} \\
Sydney, Australia}
}

\maketitle

\begin{abstract}
Modern Security Operations Centres (SOCs) integrate diverse tools, such as SIEM, IDS, and XDR systems, offering rich contextual data, including alert enrichments, flow features, and similar case histories. Yet, analysts must still manually determine which of these contextual cues are most relevant when validating specific alerts. We introduce \systemname{}, an AI assistant that learns from analysts' prior investigations to help them identify the most relevant context for new alerts. Rather than providing enrichments, \systemname{} models how analysts have previously selected context and suggests tailored cues based on the characteristics of each alert. We formulate context selection as a sequential decision-making problem and apply imitation learning (IL) to capture analysts' strategies, evaluating multiple IL approaches. Through staged evaluation, we validate \systemname{} using two intrusion detection datasets (HIKARI-2021, UNSW-NB15). In simulation-based experiments, \systemname{} helped simulated reinforcement learning analysts improve classification accuracy ($p<0.001$) (increasing F1 by 2.5\% for HIKARI and 9\% for UNSW), reducing false negatives (1.5\% for HIKARI and 10\% for UNSW), and keeping false positives below 1\%. Decision confidence among agents also improved by 2-3\% ($p<0.001$). In a within-subject user study (N=13; power = 0.8), non-experts using \systemname{} improved classification accuracy by 21.1\% ($p = 0.008$) and reduced alert validation time by 24\% ($p = 0.01$). These results demonstrate that by learning context-selection patterns from analysts, \systemname~can yield notable improvements in investigation effectiveness and efficiency. 
\end{abstract}


\section{Introduction}
\label{sec:intro}

Modern Security Operations Centres (SOCs) utilise a wide range of technologies, such as Endpoint or Extended Detection and Response (EDR/XDR), Network Detection and Response (NDR), Network Intrusion Detection and Prevention Systems (IDS/IPS), and Security Information and Event Management (SIEM), to protect modern digital enterprises. However, poor integration across security tools and the difficulty of identifying appropriate contextual information can hinder SOC analysts' ability to prioritise and respond to critical threats~\cite{Crowley2024}. 

\begin{figure*}[ht]
    \centering
    \includegraphics[width=0.85\textwidth]{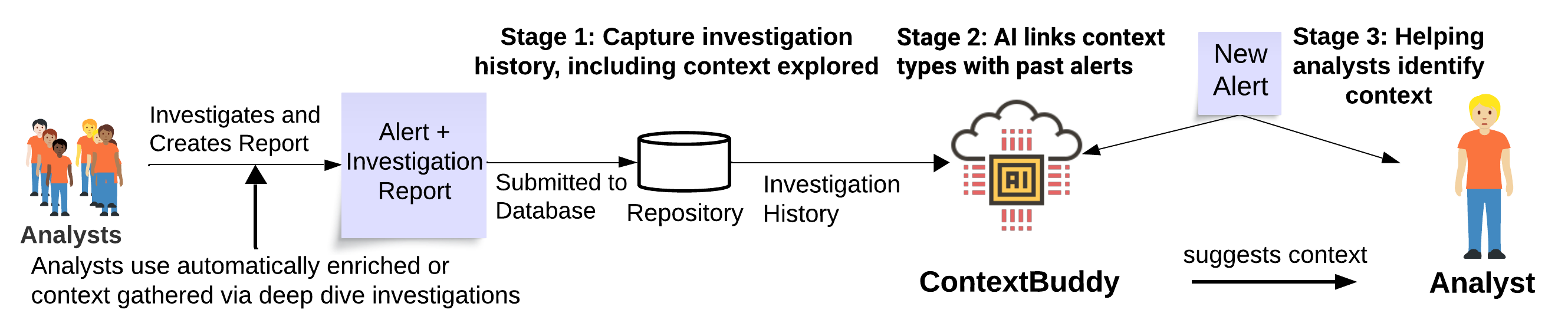}
    \caption{\systemname: an imitation learning assistant trained on prior investigation history to help analysts identify the relevant context.}
    
    \label{fig:overall-approach}
\end{figure*}

Contextual information is essential for analysts to accurately assess the severity, impact, and root causes of security alerts~\cite{Alahmadi2022-wb,Knerler2023-ja,Baruwal-Chhetri2024-yf}. A growing body of research supports methods that enhance analysts' situational awareness, such as exposing network traffic patterns and causal relationships between alerts~\cite{Dong2023-ar,Hassan2020-wc,Van_Ede2022-cp,Shen2019-ho,Ge2022-bt,Zhao2024-ma,Bhaskara2021-dq}. Additionally, AI co-pilots such as Microsoft Security Copilot~\cite{Freitas2024-sx} provide similar cases, summarise incidents, answer analyst queries, and suggest remediation steps.

As toolchains and AI-driven contexts expand, the question is no longer merely \textit{how to generate more context}, but rather \textit{how to deliver the \textbf{right} context to the analyst at the right time}. While advancing automation is important, it is not sufficient on its own. In a rapidly evolving cyber landscape~\cite{Rajivan2017-ez,Lyn_Paul2019-zw,Gomez2019-xa,Hauptman2023-zf}, effective human-AI teaming becomes increasingly critical. Analysts excel at identifying the most relevant context for specific alerts, while AI can learn from these patterns to assist them in doing so more efficiently~\cite{Seeber2020-cy,Henry2022-jv,Baruwal-Chhetri2024-yf,Tariq2024-jl,Wang2024-zz}. The human component remains vital for navigating the uncertainty and complexity inherent in cyber environments, enabling timely and accurate \textit{intrusion detection}~\cite{Shoukat2024-mt,Kim2024-jp}. Therefore, we propose reframing the task of identifying relevant context for alert or intrusion validation as a \textbf{teaming problem} that leverages the complementary strengths of human analysts and AI systems. This perspective prompts a key question:

\textit{\textbf{How can AI learn the patterns of analysts' use of contextual information from prior investigations to better guide future investigations in a human-AI team?}}

To answer this question, we propose \textbf{\systemname}, an AI assistant that learns from historical investigation data and collaborates with analysts to identify the most relevant contextual information. \systemname~operates in three stages (see Fig.~\ref{fig:overall-approach}).
In \mbox{\textbf{Stage 1}}, analysts investigate alerts using various contextual information, creating a knowledge repository that captures how different alert types benefit from particular contextual information.
In \textbf{Stage 2}, \systemname~ learns patterns linking alert characteristics to relevant contextual information. In our implementation, we primarily use imitation learning (IL). This allows \systemname~ to suggest relevant context based on features of new alerts requiring validation. 
In \textbf{Stage 3}, \systemname~ assists analysts during investigations by suggesting contextual information, either in a \textit{one-time} fashion or through an \textit{iterative} exchange of contextual information. 

To generate the data for \textbf{Stage 1}, given the scarcity of SOC-labelled data and challenges in recruiting expert analysts~\cite{Mink2023-ea,Alahmadi2022-wb,Oesch2020-wn}, we trained various reinforcement learning (RL) agents (Advantage Actor-Critic (A2C) algorithm~\cite{Mnih2016-rh}, Proximal Policy Optimisation~\cite{Schulman2017-rl}, and Deep Q-network (DQN)~\cite{Mnih2015-vz}). The RL agents validated intrusion events from two intrusion detection datasets (HIKARI-2021~\cite{Ferriyan2021-pe} and UNSW-NB15~\cite{Moustafa2015-ru}) by requesting various feature subsets (context types, like \textit{Payload Information} or \textit{Packet Counts} - see Figure~\ref{fig:cb_arch1}) to generate investigation traces consisting of which contexts the RL agents requested and their final decision. In \textbf{Stage 2}, we trained \systemname{} using various IL methods (Adversarial Inverse Reinforcement Learning~\cite{Fu2017-rc}, Generative Adversarial Imitation Learning~\cite{Ho2016-uc}, and Behaviour Cloning~\cite{Torabi2018-ez}), finding that GAIL was most effective at learning analysts' context usage patterns.

In \textbf{Stage 3}, we evaluated \systemname{} in two experiments: simulation-based by teaming \systemname{} with the RL agents it was trained on, and a follow-up user study with non-experts. Our results demonstrate the effectiveness of \systemname{} in helping simulated analysts or users improve decision accuracy, confidence, and reduce false negatives and positives. Most of these gains were statistically significant. While the results in \textit{iterative} mode were mixed, \textit{one-time} mode resulted in the highest improvements.

\subsection*{Contributions}

This paper makes the following contributions:
\begin{enumerate}[leftmargin=*]
    \item \textbf{Introduction of \systemname:} 
    We introduce \systemname, a framework leveraging IL to learn analysts' patterns of context use, and suggest relevant contextual information with explainability to enhance the accuracy and efficiency of alert investigations.
    \item \textbf{Interaction Modes:} 
    We examine two modes for communicating context to analysts: \textit{One-time} (one-off context suggestion) 
    and \textit{Iterative} (multiple rounds; tested in user study only), 
    and analyse their relative effectiveness.
    \item \textbf{Demonstrating Effectiveness:} 
    We demonstrate the effectiveness of IL approaches in capturing diverse analyst strategies through rigorous simulation-based validation.  Additionally, in a user study with non-experts (non-SOC analysts), the \textit{One-time} mode boosted their classification accuracy by 21.1\% and reduced task time by 24\%, indicating faster investigations. 
\end{enumerate}

\section{Background}
\label{sec:background}

SOC analysts face numerous challenges including dealing with alert overload and lack of contextual information, leading to analyst fatigue and burnout \cite{Alahmadi2022-wb}. While machine learning has shown promise in automating various tasks~\cite{Arp2022-ec}, including intrusion detection~\cite{Ferrag2020-fu,Han2020-ui,Hassan2020-wc}, malware analysis~\cite{Saha2024-gb}, and threat intelligence~\cite{Mallikarjunaradhya2023-xw}, relying solely on machine intelligence can result in missed detections due to inaccuracies inherent in machine learning models~\cite{Wang2024-zz,Kim2024-jp,Shoukat2024-mt}. Many works correctly advocate for human involvement in intrusion detection to complement automated systems, reduce false positives, leverage human judgment, and ultimately improve the accuracy and timeliness of threat detection~\cite{Shoukat2024-mt,Kim2024-jp,Baruwal-Chhetri2024-yf}.

\subsection{Contextual Information for Improving Decision-Making}
\textit{Context} provides analysts with an understanding of an alert's potential implications~\cite{Knerler2023-ja,Kovacevic2023-fi}. Several studies have explored ways to extract contextual information from raw data.  UNICORN~\cite{Han2020-ui} use provenance graphs to \textit{trace critical system activities}, while ATTACK2VEC~\cite{Shen2019-ho} models the \textit{evolution of cyberattacks} through temporal embeddings. DEEPCASE~\cite{Van_Ede2022-cp} identifies \textit{preceding events to the alert}, and Context2Vector~\cite{Liu2022-xl} uses representation learning to transform raw events into \textit{event sequences (source, target, and tuple)}.
RAPID~\cite{Liu2022-af} captures \textit{the causal connection between different alerts}, and DrSec~\cite{Sharif2024-bt} employs self-supervised learning to enrich alerts with data, such as \textit{system configurations and MITRE ATT\&CK references}. 

Existing research primarily emphasises automated detection techniques, with less attention given to human-AI collaboration and the specific needs of analysts. MStream~\cite{Bhatia2021-oy} detects anomalies in multi-aspect data streams by deriving contextual information from, e.g., \textit{IP addresses and packet sizes} to group anomalies. Trident~\cite{Zhao2024-ma} enhances attack detection through incremental learning to capture network traffic patterns (such as \textit{protocol flags and behaviours}) for specific attack types. Net-track~\cite{Lee2023-eq} detects web trackers using packet metadata. Others use transformer-based approach~\cite{Hang2023-dt,Manocchio2024-bg}. 

While tools like AI co-pilots~\cite{Freitas2024-sx} support individual analysts during investigations, \systemname{} aims to complement such systems by learning a generalised context-selection policy from historical traces across multiple analysts. We can construct a rich, multi-analyst corpus of sequential context-selection behaviour by instrumenting co-pilot sessions, e.g., logging which context sources are reviewed and when. Trained on this data, \systemname{} would distil collective patterns that support consistent, team-level decision-making and accelerate context access in future alerts.

While the above works focus on technical improvements or automating detection, we aim to learn how analysts leverage diverse contextual cues to inform their decision-making in a human-AI teaming setting. The human component is necessary to manage inherent uncertainty and complexity in intrusion detection~\cite{Shoukat2024-mt,Kim2024-jp}. We provide a pathway to realise this by employing IL to capture analysts' tacit knowledge; \systemname~ learns to understand and apply contextual information analysts find relevant, enabling context-sensitive alert investigations.

\subsection{AI-Assisted Decision-Making}

To help SOC analysts understand the relevance of \systemname's feature selections for each alert, it is crucial to provide explainability information to analysts. The types of information \systemname~should provide is an important consideration for effective human-AI teaming. When AI offers \textit{recommendations} and explanations, and humans review them, it can expedite decision-making but may lead to over-reliance on AI~\cite{Fogliato2022-dj, Bucinca2021-di}. Conversely, cognitive forcing techniques, like delaying recommendations, can mitigate over-reliance but may inadvertently cause under-reliance, particularly among experienced analysts who might distrust AI~\cite{Fogliato2022-dj,Gajos2022-is,Bucinca2021-di}. 

A new approach, \textit{evaluative AI}, focuses on supporting human decision-making by providing evidence for and against the event types rather than recommending a specific classification~\cite{Miller2023-uz}. Integrating evidence-based Explainable AI (XAI) without recommendations (\textit{machine-in-the-loop framework}) can enhance decision-making~\cite{Le2024-mc,Vered2023-wo} by empowering users to take control~\cite{Shneiderman2022-uj}. Our user study employs evidence-based XAI, presenting evidence for and against each intrusion class within a machine-in-the-loop framework that refrains from recommending a specific class.

\subsection{XAI for IDS}
We opted for SHAP (SHapley Additive exPlanations)~\cite{Lundberg2017-ns} as our evidence-based XAI approach. Based on Shapley values from game theory~\cite{Shapley1953-kj}, SHAP is an XAI method that interprets machine learning models by assigning importance scores to features. SHAP quantifies how much each feature influences the model's output. SHAP remains a leading XAI approach for intrusion detection due to its high accuracy and interpretability~\cite{Abdulganiyu2023-ch, Nwakanma2023-mx,Durojaye2024-pr, Corea2024-td, Arreche2024-nd}. An alternative approach is to use AE-pvalues, a post-hoc explanation technique using Autoencoders for explaining intrusion events~\cite{Lanvin2023-ts}. Despite concerns about computational complexity, SHAP's interpretability and trustworthiness make it ideal for understanding AI decisions~\cite {Barnard2022-pf, Tritscher2023-ge}. Thus, our work uses SHAP-based XAI.

\begin{figure*}[ht]
    \centering
    \includegraphics[width=\textwidth]{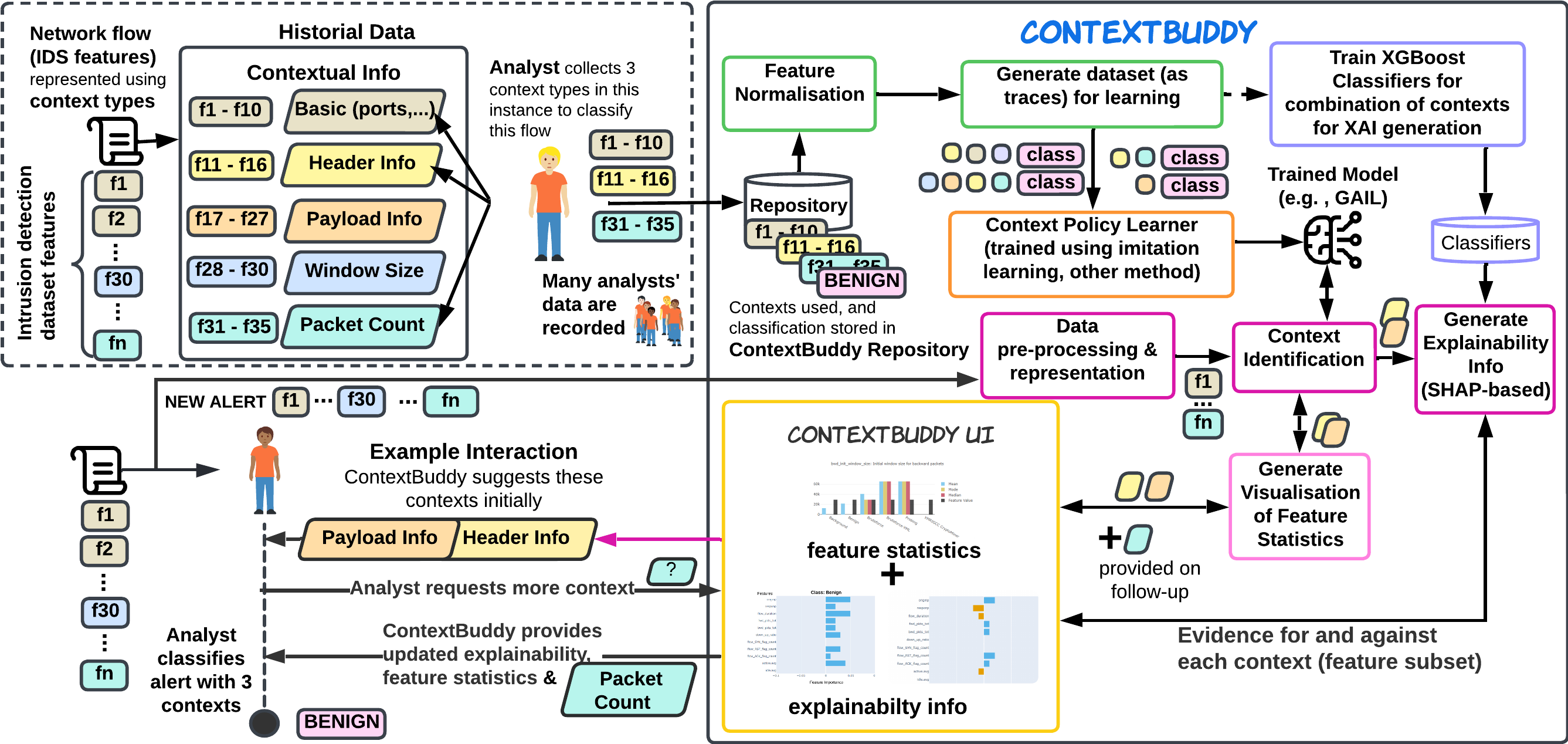}
    \caption{Main components of \systemname{} and an example of a user's interaction with it.}
    \label{fig:cb_arch1}
\end{figure*}

\section{Proposed ContextBuddy for Contextual Information}
This section introduces \systemname, its components and the two interaction modes available to end users.

\subsection{\systemname~Model}
We model analysts' decision-making using Markov Decision Processes (MDPs)~\cite{Puterman2014-hc}, a framework for sequential decision-making. The analyst model is defined by the tuple $(\mathcal{S}, \mathcal{A}, T, R, \pi, \gamma)$, where $\mathcal{S}$ denotes a finite set of states representing the environment's possible configurations, $\mathcal{A}$ represents a finite set of actions available to the agent, $T: \mathcal{S} \times \mathcal{A} \times \mathcal{S} \rightarrow [0, 1]$ is the deterministic transition function indicating the next state for each state-action pair, $R: \mathcal{S} \times \mathcal{A} \times \mathcal{S} \rightarrow \mathbb{R}$ is the reward function mapping each transition to a scalar reward, and $\gamma \in (0, 1]$ is the discount factor prioritising immediate rewards over distant ones. The analyst aims to maximise expected return by optimising policy $\pi$, learning the optimal action sequence for the highest cumulative reward over time.

The actions correspond to acquiring different contextual information sequentially, each request being a distinct action. Each state represents the contextual information analysts hold. Given the sequence of actions (demonstrations showing the contextual information requested by analysts), we train \systemname~ using IL to mimic analyst behaviour from demonstrations~\cite{Zare2023-mg,Zheng2022-rp,Ho2016-uc}. However, we can switch to any appropriate method from IL.

\subsection{System Architecture of \systemname{}}

Figure~\ref{fig:cb_arch1} outlines the architecture of \systemname{}. These modules enable \systemname{} to learn from past investigations and offer scenario-specific, explainable context suggestions that support analysts during alert triage.

\begin{itemize}

    \item \textbf{\systemname{} Repository} Stores structured representations of alert investigations, including alert metadata, selected contexts, and outcomes. It forms the historical knowledge base from which the policy learner is trained.
    
    \item \textbf{Normalisation Module} performs standard preprocessing tasks on alert features from analyst investigation histories. It parses selected context types (e.g., packet counts, payload information) and formats each investigation instance for IL. This includes generating structured feature subsets for training context-specific classifiers.
    
    \item \textbf{Context Policy Learner} Trains a context-selection policy using IL (e.g., GAIL) to replicate the context-selection patterns of past analysts. 

    \item \textbf{Classifier Store (Classifiers)} Maintains a library of pre-trained classifiers (e.g., XGBoost models), each trained on a distinct subset of features corresponding to specific combinations of context types to assist in generating Shapley-based explainability information.

    \item \textbf{Context Identification} Given a new alert, \systemname~uses this module to generate a list of possible contexts and passes these to the other modules to generate feature statistics and explainability information.
    
    \item \textbf{ContextBuddy Interface (CB UI)} Provides an interface for interaction (\textit{one-time} or \textit{iterative}). It visualises explainability information and feature statistics that allow analysts to inspect underlying data. When the user requests different contexts, \textbf{Context Identification} updates the context list, feature statistics and explanations.
    
\end{itemize}

\subsection{Extracting Insights from Analyst behaviour} \hfill

\systemname~collects demonstrations of analyst behaviour to define an imitation policy $\pi^*$ of the analysts' policy $\pi$. A demonstration $\tau$ is a sequence of state-action pairs:
\[
\tau = \{(s_0, a_0, r_0),\ (s_1, a_1, r_1),\ \ldots\},
\]
where actions $a_t \in \mathcal{A}$ are requests for contextual information (e.g., packet counts, payload data), states $s_t$ include initial features and the contextual information obtained up to time $t$, and the analyst rewards $r_t$ is computed based on (see Sec~\ref{sec:simulation-study-analyst-model}) prediction accuracy (positive for correct and negative for incorrect classification), the confidence in the prediction, and the number of contextual information used. States include the initial information provided to the analyst ($f_0, \ldots, f_n$), augmented with binary encoding $c_k$, indicating whether each type of contextual information has been requested. In addition, a demonstration comprises floating-point rewards and a terminal variable capturing analysts' classifications. 

Given a set of analysts $\mathcal{J}$, \systemname~collects trajectories $\tau_j^e \in \mathcal{T}$ from each analyst $j \in \mathcal{J}$ for each alert $e \in \mathcal{E}_j$, where $\mathcal{E}_j$ represents alerts investigated by analyst $j$. These demonstrations capture the contextual information explored by each analyst. Using $\mathcal{T}$, \systemname~ learns the policy $\pi^*$. We assume that analysts behave rationally, aiming to maximise expected rewards, the environment transitions are deterministic, and the set of possible contextual information types is finite and known.

\subsection{Modes of Interaction} 
\systemname{} supports following interaction modes: 

\subsubsection{One-time Assistance}
In this mode, \systemname~proactively identifies potentially relevant contextual information for an alert and suggests it to an analyst upfront. This information is chosen based on \systemname's policy $\pi^*$, which aims to suggest an optimal sequence of actions, i.e., the contextual information that could lead to a correct decision with low entropy. It informs the analyst if no solution is found. This process aims to augment the analyst's decision-making capabilities by identifying and providing relevant contextual information that the analyst might initially overlook. This mode resembles \textit{augmentation}~\cite{Tariq2024-jl}, where AI suggests context in a one-off interaction.

\systemname~generates a plan, $P = \{a_1, a_2, ..., a_k\}$ (sequence of actions), where $P \in \mathcal{P}$ is a plan from the set of all possible plans, $\mathcal{P}$, to maximise the expected reward based on its policy ($\pi^*$):
\[
P = \underset{P \in \mathcal{P}}{\mathrm{argmax}}\, R(P)
\]
where $R(P)$ represents the expected reward over the plan $P$. 

\subsubsection{Iterative Assistance}
This mode fosters a collaboration between the analyst and \systemname. It is a multi-step process with an iterative exchange of contextual information that continues until the analyst feels confident enough to classify the intrusion alert~\cite{Tariq2024-jl}. Either party could initiate the process.  

\systemname~leverages its policy ($\pi^*$) and the history $H$ of its and the analyst's previous selections to determine one or more contextual information to suggest. Examining the history helps it understand past selections and anticipate potential outcomes. \systemname~ selects the action $a_i^{t+1}$ that maximises the expected reward over prior actions:
\[
a_i^{t+1} = \underset{a \in A \setminus H}{\mathrm{argmax}}\, R_i(a|H)
\]

$R_i(a|H)$ is the expected reward for taking the next action $a$ after the prior action sequence $H$, and action $a_i^{t+1}$ maximises this reward.

\section{\systemname{}~Evaluation}
\label{sec:study-design}

To show whether \systemname{} could effectively learn from analyst demonstrations and improve analysts' performance by teaming up with them, we conducted a simulation and a user study to evaluate \systemname. This section discusses the analyst's task of categorising intrusion features to form context categories that analysts request. 

\subsection{Investigation Task}

We evaluated \systemname{} using a multi-class intrusion classification task where analysts (simulated or humans) classify intrusion events into one of many classes by requesting and analysing contextual information (feature subsets). We utilised the HIKARI-2021~\cite{Ferriyan2021-pe} and UNSWNB-15~\cite{Moustafa2015-ru} open-source intrusion detection datasets, excluding source and destination IP addresses to prevent model bias towards specific IPs. We used all other features: 83 for HIKARI, and 47 for UNSWNB-15. The datasets provide diverse attack coverage and encrypted traffic, with HIKARI focusing on web-based application-layer attacks.

Previous research has identified five complementary types of context for intrusion detection~\cite{Aleroud2017-ow}: \textit{Individuality Context}, capturing network entity characteristics (e.g., device type, known vulnerabilities); \textit{Activity Context}, recording network actions such as logs and user operations; \textit{Location Context}, encompassing geographical or network‐topological data; \textit{Time Context}, reflecting temporal patterns and intervals; and \textit{Relation Context}, describing interdependencies among events and entities. Furthermore, building on various works~\cite{Zhou2024-iw,Shanker2023-gr,He2024-tp} on the idea of \textit{semantic feature grouping} and taking inspiration from these works, we designed our contextual information categories for the two intrusion datasets. For example, He \emph{et al.} grouped raw IoT traffic fields into protocol‑semantic blocks (e.g., basic, flags, payload), delivering lightweight yet context‑rich features for attack detection~\cite{He2024-tp}. In our paper, this organisation (grouping features into categories for intrusion detection) is aimed at mirroring the real‐world evidence integration process in SOC environments (e.g., SOAR platforms~\cite{Muniz2021-ia}), enabling simulations in which analysts iteratively request and combine contextual information in a manner that balances investigative realism when we do not have real-world SOC analysts' data.

\section{Evaluation Setup (Simulation-based)}
\label{sec:study-design-sim}

This section discusses the design of our first simulation-based study. In the absence of an analyst-annotated dataset, we generated synthetic data to train \systemname. We employed RL to train simulated analysts to classify intrusion events and then trained \systemname{} using various IL methods on the simulated analysts' decisions. To test \systemname's effectiveness at helping these analysts improve their performance, we then tested \systemname{} by teaming it with the same simulated analysts on a new set of events.

\subsection{Independent Variables:}
We evaluated simulated analyst performance with and without \systemname's assistance.

\begin{itemize}[leftmargin=0.5cm]
    \item \textit{Analyst (Baseline):} The analyst works independently. These are our RL (A2C, PPO, and DQN) baselines.
    \item \textit{Assistant:} \systemname{} works autonomously. This is another baseline when comparing the two teaming conditions.
    \item \textit{One-time Assistance:} \systemname{}  provides one-off contextual suggestions to the analyst.
\end{itemize}

\subsection{Dependent Variables:}
\begin{itemize}[leftmargin=0.5cm]
    \item \textbf{Classification Accuracy:} A correct classification aligns with the ground truth recorded in the dataset.     
    \item \textbf{Confidence:} We measure how confident the simulated analyst is in its prediction.
     
\end{itemize}

\subsection{Dataset Preparation}
We selected 16,000 alerts (32,000 across both datasets), ensuring the selection represents the underlying class distributions (see Appendix A (Figs. 10 \& 11) for distribution by attack type). These were split into 15,000 historical alerts and 1,000 new alerts per dataset. We divided the historical and new alerts into 10 subsets to avoid results being biased by specific instances. When we split the 16,000 instances into 10 subsets, each contained 1500 randomly selected historical and 100 new alerts. To introduce variability in analyst behaviour, we trained three analysts for each subset, each receiving 500 historical alerts and tested against the same 100 new alerts. This number allowed the analysts to achieve a reasonable level of classification accuracy without extensive training, effectively representing trained, high-performing analysts. While providing more alerts per analyst could improve training, we focused on achieving realistic classification accuracy and efficiency without requiring extensive training time.

\subsection{Analyst Model}
\label{sec:simulation-study-analyst-model}

To generate synthetic traces, we trained various RL models as agents. The analyst's investigation process is as follows (see Fig. 25 in appendix more details):
\begin{enumerate}[leftmargin=0.5cm]
    \item \textbf{Alert Assignment:} The analyst is assigned a random alert and initial features not included in contextual categories.
    
    \item \textbf{Requesting Context:} The analyst requests specific contextual information (e.g., \textit{Payload Information}) and retrieves the relevant feature subset (all payload-related features).
    
    \item \textbf{Training Classifier:} The analyst integrates new features and trains an XGBoost classifier on $120,000$ balanced instances (with the same number of samples of each class/attack type), applying oversampling of less common classes, a common method for tackling imbalanced datasets~\cite{Haixiang2017-so}.
    
    \item \textbf{Classification:} The analyst requests contextual information (feature subsets) until it is ready to classify an alert. The analyst may request all available contextual information or only a subset when classifying each alert. After each classification, the RL agents get an end-of-episode reward (explained below). 
    
    \item \textbf{Iteration:} The analyst repeats this process to classify each of the 500 alerts. Since we train an agent for a higher number of time steps, the agent is exposed to the same alert multiple times to train it to identify the best possible set of contextual information for classifying different alerts.
\end{enumerate}

The \textit{observation space} included: 1) the feature subsets (feature values) collected by the analysts; 2) a set of variables, one for each type of contextual information, to track which contexts it has collected; this is represented by a state variable that ranges from 0 to 2, with 0: not requested, 1: requested once, 2: requested 2 or more times); this range assists the agent in distinguishing between states where it requested a specific context multiple times; 3) its prediction confidence; and 4) the ratio of repeated actions and total actions (ideally, should be zero). This observation space captures the individual features and the requested contexts and allows the agent to associate these requests with their confidence level and whether they requested unique information. The \textit{action space} included an integer from 0 to the number of possible contexts $+1$; this additional action represents the \texttt{classify\_alert} action. 

We shape the simulated analyst's reward to encourage accurate and confident decisions while promoting efficient use of context. The base classification reward reflects both correctness and confidence:
\begin{align}
\text{reward}_{\text{classify}} =\;&
\bigl(\text{correct\_reward} + \text{conf}\bigr)\ \times I(\text{predicted} = \text{true}) \notag \\\notag 
&+ \bigl(\text{incorrect\_penalty} - \text{conf}\bigr)\ \\  &\times I(\text{predicted} \neq \text{true})\\
&+ \phi \cdot I(\text{correct} \land \text{high\_conf}) \notag \\
&+ \psi \cdot I(\text{correct} \land \neg\text{high\_conf}) \notag \\
&- \omega \cdot I(\text{no\_context\_used})
\end{align}

\noindent
To guide learning, we incorporate reward shaping terms that encourage strategic use of context:
\begin{align}
\text{reward}_{step} =\;& \lambda_1 \cdot \text{num\_requests} -\; \lambda_2 \cdot \text{num\_repeats} \notag \\
&+ \eta_1 \cdot \text{num\_novel}
\;+\; \eta_2 \cdot \max(0, \Delta\text{conf})
\end{align}
\noindent where:
\begin{itemize}[leftmargin=1.5em]
    \item $\lambda_1$: penalty per context request (-0.02),
    \item $\lambda_2$: penalty for repeated requests (-0.5),
    \item $\eta_1$: bonus for novel (new) context requests (0.2),
    \item $\eta_2$: bonus for increased prediction confidence (difference in confidence),
    \item $\phi$: reward for correct and confident classifications (+10),
    \item $\psi$: reward for correct but less confident classifications (+5),
    \item $\omega$: penalty for classifying without using any context (-5).
\end{itemize}

Each classification grants a base accuracy reward plus shaping bonuses or penalties that push the analyst to request just enough context, guiding them on what to ask for and when to stop and decide.

\subsection{Analyst Model Training}
\label{sec:study-analystmodel-reward}
Analysts were trained using three RL methods: Advantage Actor-Critic (A2C) algorithm~\cite{Mnih2016-rh}, Proximal Policy Optimisation~\cite{Schulman2017-rl}, and Deep Q-network (DQN)~\cite{Mnih2015-vz}. We used Optuna\footnote{\url{https://optuna.org/}} over 2000 trials to automate hyperparameter search for $\gamma$ and $ent\_coeff$, leaving others to default values in StableBaselines3. Our model's parameters are shown in Table~\ref{tab:hyperparams-rl}, and we cross-checked that these are typical values used across various environments tested by RL Baselines3 Zoo\footnote{\url{https://github.com/DLR-RM/rl-baselines3-zoo/tree/master/hyperparams}}. Each model was trained for a maximum of $300,000$ time steps or if a predefined positive reward threshold is reached (implemented via \textit{StopTrainingOnRewardThreshold} callback\footnote{\url{https://stable-baselines3.readthedocs.io/en/master/guide/callbacks.html}}), consistent across all analysts and data subsets. A2C used RMSprop as its optimiser (default in SB3), whereas PPO and DQN used Adam. All experiments were conducted on a high-performance computing (HPC) system equipped with NVIDIA H100 GPUs and CUDA 11 support. Each analyst model was trained sequentially within its subset using 1 NVIDIA H100 GPU and 32 GB of RAM. However, training across the 10 subsets was parallelised to maximise GPU utilisation across the HPC environment.

\begin{table}[h]
    \centering
    \begin{tabular}{lcccp{3cm}}
        \toprule
        Algorithm & $\gamma$ & ent\_coef & Policy & Policy Architecture \\
        \midrule
        PPO  & 0.99 & 0.01  & MlpPolicy  & NN with 2 hidden (64 * 64) (ReLU) \\
        A2C  & 0.99 & 0.001 & MlpPolicy   & same as above \\
        DQN & 0.99 & 0.12 & MlpPolicy  & same as above\\
        \bottomrule
    \end{tabular}
    \caption{Hyperparameters for RL.}
    \label{tab:hyperparams-rl}
\end{table}

\subsection{\systemname~Training}

We trained \systemname{} using AIRL~\cite{Fu2017-rc}, GAIL~\cite{Ho2016-uc}, and BC~\cite{Torabi2018-ez}; we used implementations from an existing IL library\footnote{\url{https://github.com/HumanCompatibleAI/imitation}}. For each data subset, we trained three IL agents (AIRL, GAIL, and BC) using 1 NVIDIA H100 GPU and 64 GB of RAM. Training across the 10 subsets was parallelised.

\begin{itemize}[leftmargin=0.5cm]
    \item 
        \textbf{Data Collection:} \systemname{} collected state-action trajectories from three analysts, each investigating 100 alerts, resulting in 300 trajectories per subset. When training \systemname{} with input from all models (A2C, PPO, and DQN), we used 100 trajectories from each model. 

    \item 
        \textbf{Model Components:} The reward network used by the discriminator was \texttt{BasicShapedRewardNet}, a potential function network with two hidden layers of 32 units each for modelling potential-based reward shaping. We configured the A2C algorithm learner or policy model using the default MlpPolicy architecture described above. The replay buffer was set to 3,000, and the discriminator was updated 10 times per training round.

    \item 
        \textbf{Training Process:} Training alternated between the generator and discriminator until the number of transitions sampled exceeded 200,000 timesteps. With approximately 2,700 transitions generated by 300 alerts (each resolved in an average of 9 actions), 200,000 timesteps ensured sufficient sampling and training. We experimented with 50,000 to 300,000 timestamps and noted no significant improvement beyond 200,000 timesteps.
\end{itemize}

\section{Evaluation Results  (Simulation-based)}
\label{sec:results}

In this section, we report the simulation study's results. We first discuss the performance of individual analysts, then \systemname{} against analysts, and finally the results of the teamwork between them. We conducted pairwise comparisons using Wilcoxon signed-rank tests with Bonferroni corrections as required, using Cohen's $d$ for effect size. When dealing with binary measures (accuracy), we used McNemar's test with Cohen's G effect size measure.

To ensure robustness and generalisability, we had constructed 10 stratified datasets across \texttt{HIKARI} and \texttt{UNSW}, each with 1,500 historical and 100 new alerts reflecting original class distributions, and trained three analyst models per dataset on 500 alerts each. When reporting results, we aggregate performance across all 10 subsets. Furthermore, whenever results are presented for a given RL model (e.g., pairwise comparisons), they reflect the collapsed performance across the \textbf{three simulated analysts} trained per subset. This aggregation was necessary to capture the average behaviour of the RL models under diverse but matched training conditions, thereby enhancing the external validity of our comparisons. 

\begin{figure}[h]
    \centering
    \includegraphics[width=0.85\linewidth]{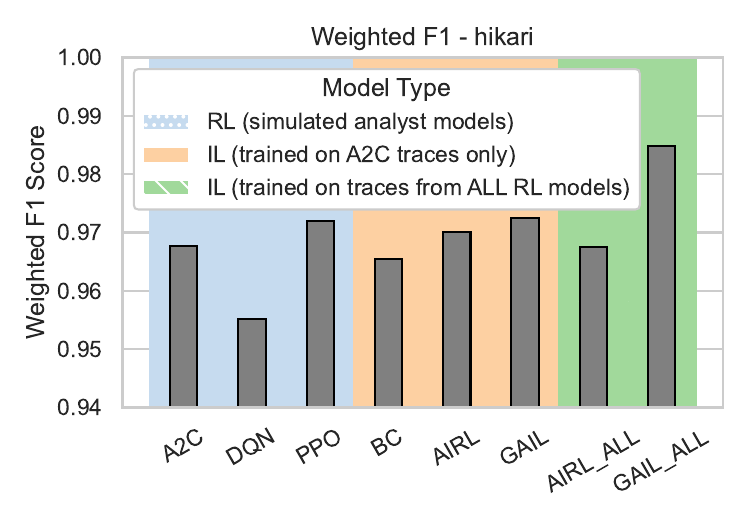}
    \caption{Comparing F1 scores across RL and IL models for HIKARI dataset.}
    \label{fig:box_f1_HIKARI}
\end{figure}

\begin{figure}[h]
    \centering
    \includegraphics[width=0.85\linewidth]{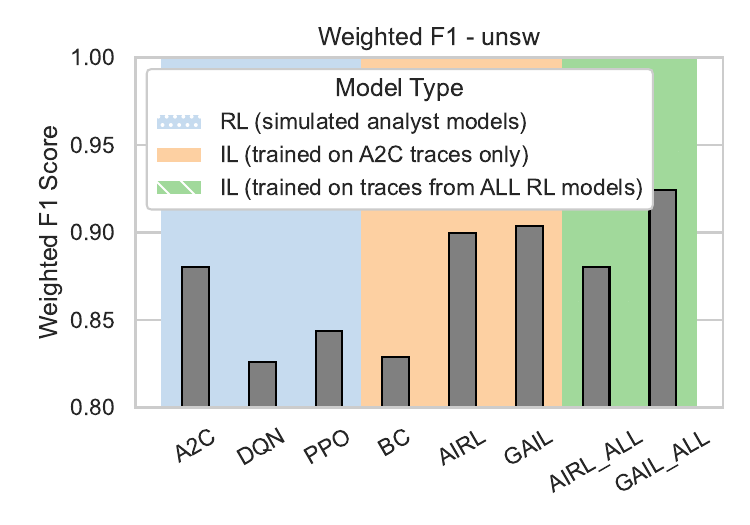}
    \caption{Comparing F1 scores across RL and IL models for UNSW dataset.}
    \label{fig:box_f1_UNSW}
\end{figure}

\subsection{\textbf{Performance of Simulated Analyst (RL) Models}} 

\textbf{Across both domains, A2C and PPO were best performers}, with only one pairwise comparison, \texttt{A2C} outperforming \texttt{DQN} in \texttt{UNSW} was statistically significant ($p = .0008$, $g = 4.84$, \textit{large effect}). For the HIKARI domain, the \texttt{PPO} model achieved marginally higher correct predictions relative than other two, but these differences did not reach significance after adjustment. For the UNSW, no significant difference emerged between other comparisons. 
\vspace{5pt}

\subsection{\textbf{Performance of \systemname{} (IL) vs Simulated Analysts}} 
We trained IL methods on traces from A2C only (given the result from the UNSW dataset above, and on traces from all 3 models (\texttt{GAIL\_ALL} and \texttt{AIRL\_ALL}). We conducted pairwise performance comparisons to assess whether IL models effectively learned from simulated analysts.  

\textbf{In HIKARI}, most comparisons showed no statistically significant difference. We note that any improvements in single-source IL methods were driven by the fact that they were trained on three instances of A2C models, each trained on a different set of alerts, providing the IL method with traces from different analysts but trained using A2C. \textbf{Notably, \texttt{GAIL\_ALL} consistently outperformed all others}, implying a higher capacity to extract and generalise useful strategies from observing multiple RL models. \texttt{GAIL\_ALL} demonstrated a statistically significant advantage over \texttt{A2C} ($p_{\text{adj}} = 0.0021$, $g = 0.5$) and \texttt{DQN} ($p_{\text{adj}} < 0.001$, $g = 0.96$), indicating strong evidence of successful learning, and 1.3\% increase over \texttt{PPO}. 

\textbf{In UNSW}, the results were more definitive, with numerous statistically significant differences. Several IL models, particularly \textbf{\texttt{GAIL} and \texttt{GAIL\_ALL}, outperformed their RL counterparts}. For example,  \texttt{GAIL\_ALL} significantly outperformed \texttt{DQN} ($p_{\text{adj}} < 0.001$, $g = 1.0$). These findings indicate that in the \texttt{UNSW} domain, IL models, particularly \texttt{GAIL\_ALL}, effectively internalised and, in some cases, exceeded the decision-making strategies demonstrated by the RL analysts. 

\textbf{Our results demonstrate that IL can be a viable path to replicating and even enhancing the decision strategies of simulated analysts}. Later experiments in the paper were conducted using \systemname{} trained on \texttt{GAIL\_ALL}.

IL models, particularly \textbf{\texttt{GAIL\_ALL}, demonstrated strong performance in detecting threats by maximising true positives (\textit{TP}) and reducing false negatives (\textit{FN})}. In the \texttt{UNSW} domain, \texttt{GAIL\_ALL} reduced \textit{FN} by over 40\% relative to \texttt{DQN} ($p < 0.001$), with similar trends observed against \texttt{PPO} and \texttt{A2C}. However, this improved recall came with increased false positives (\textit{FP}), especially in single-source models like \texttt{GAIL} and \texttt{AIRL}. \texttt{GAIL\_ALL} improved FP coverage further, with \texttt{GAIL\_ALL} showing a 10 - 15\% drop relative to  \texttt{PPO} in \texttt{UNSW} ($p < 0.01$, $r \approx 0.35$). 

\textbf{Regarding predictive confidence, \texttt{GAIL\_ALL} significantly outperformed all RL baselines}. Against \texttt{A2C}, it achieved higher median confidence ($0.9933$ vs.\ $0.9910$, $p < 0.001$, $r = 0.32$), \texttt{DQN} ($0.9932$ vs.\ $0.9834$, $p < 0.001$, $r = 0.41$) and \texttt{PPO} ($0.9931$ vs.\ $0.9916$, $p < 0.001$, $r = 0.29$). 

While \systemname{} tackles human–AI teaming problem that lacks established baselines, Transformer- and LLM-based models offer a natural point of reference for comparing \systemname's individual performance due to their growing use in IDS classification tasks. Recent works have applied these approaches to the UNSW dataset, with reported F1 scores ranging widely, 0.74~\cite{abdennebi2024machine}, 0.59~\cite{zhao2025efficient}, 0.56~\cite{alzahrani2022anomaly}.  
This suggests that \systemname{} is performing multi-class classification better and is also in a position to offer users context suggestions. 

\subsection{\textbf{\systemname{} and Analysts Team Performance}}

To investigate how analysts should collaborate with \systemname{}, we designed a set of \textit{suggestion adoption strategies} that vary in when and how analysts consider additional context cues suggested by \systemname{}. This design enables a controlled and systematic evaluation of teaming strategies and their impact on performance.

Each simulated analyst's policy outputs a sequence of context features for each alert. These contexts determine the subset of features provided to a pre-trained XGBoost classifier, producing a class prediction. Each XGBoost model is trained offline on a specific context subset; analysts reuse these pre-trained models without retraining during training and testing. The classifier's softmax-normalised probability for the predicted class is used as a scalar confidence score. The analyst follows a two-stage process when deciding whether to accept the suggested context:

\begin{enumerate}
    \item The analyst executes its derived plan (sequence of context features) and obtains a prediction and confidence score.
    \item It then evaluates any \textit{additional, non-overlapping} contexts suggested by \systemname{}. If the prediction using the extended context leads to higher confidence, the analyst accepts the new decision; otherwise, it retains its original decision. This setup models a realistic constraint: analysts cannot ``unsee'' it once the context is collected. 
\end{enumerate}

\textbf{Teaming Strategies.}
We tested the following analyst-AI collaboration strategies:

\begin{itemize}
    \item \textbf{Alone (no AI support):} analyst ignores all suggestions and relies solely on their own RL policy.
    \item \textbf{Always Consider (no threshold):} Analyst always evaluates additional context suggested by \systemname{}.
    \item \textbf{Random:} Analyst accepts suggestions at random with 50\% probability. A fixed seed ensures reproducibility.
    \item \textbf{Threshold-based:} Analyst evaluates \systemname's suggestion only if its confidence is below a specified threshold (0.90, 0.80, 0.70, or 0.60). 
\end{itemize}

These strategies are applied uniformly across all domains and models. Analysts adopt the AI-suggested context only if the resulting prediction is more confident. Even if the AI suggestion results in a correct prediction with lower confidence, the analyst defaults to the original decision. 

\begin{figure}[h]
    \centering
    \includegraphics[width=\linewidth]{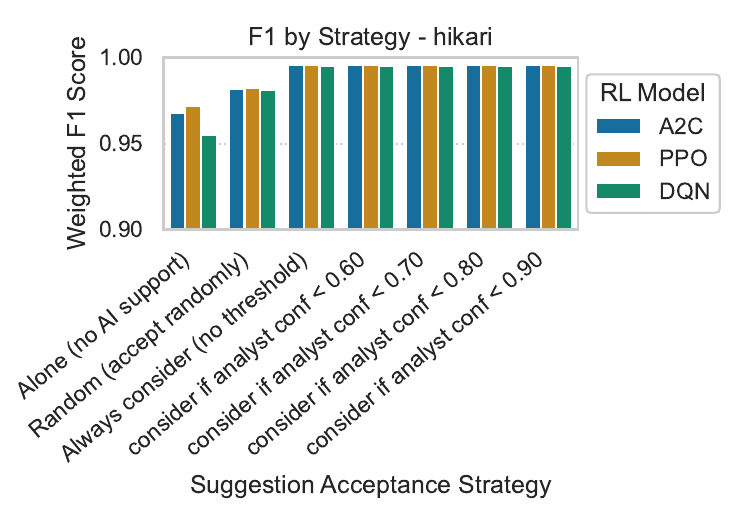}
    \caption{The performance gain when simulated analysts team up with ~\systemname in HIKARI dataset.}
    \label{fig:f1_teaming_barplot_HIKARI}
\end{figure}

\begin{figure}[h]
    \centering
    \includegraphics[width=\linewidth]{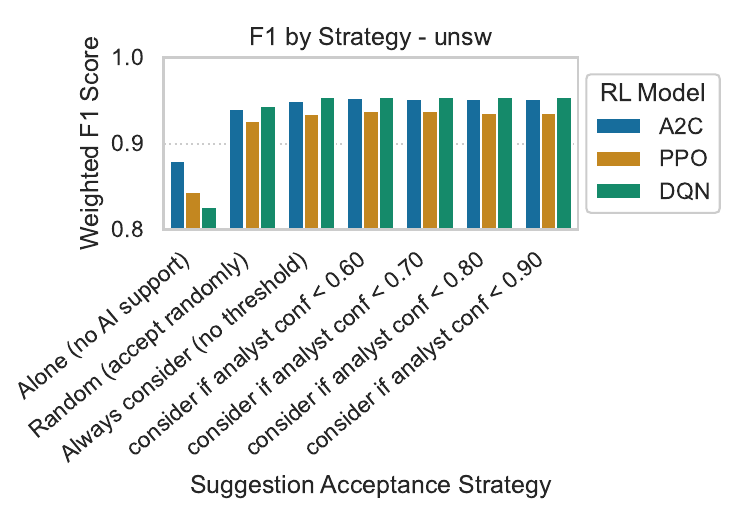}
    \caption{The performance gain when simulated analysts team up with ~\systemname in UNSW dataset.}
    \label{fig:f1_teaming_barplot_UNSW}
\end{figure}

We show the results for the two domains in Figure~\ref{fig:f1_teaming_barplot_HIKARI} and Figure~\ref{fig:f1_teaming_barplot_UNSW} for HIKARI and UNSW, respectively. The \textbf{Threshold-based} strategy demonstrates a significant improvement over the \textbf{Working Alone} baseline ($p<0.001$, with large effects; Cohen's $g >= 0.4$) across both datasets. The \textbf{Random} baseline yielded modest but non-trivial accuracy gains, with Cohen's $g$ ranging from 0.20 to 0.50 and several $p$-values indicating significance ($p < 0.001$), albeit with greater variability and risk of incorrect adoption. Across all models in the \textit{HIKARI} domain, McNemar tests yielded highly significant results (all $p < 0.0001$) as threshold levels decreased (i.e., as models increasingly accepted \systemname's suggestion). A parallel pattern was observed in the \textit{UNSW} domain. 

Regarding the simulated analysts' confidence (Figs. 32 \& 33 in appendix), the \textbf{no threshold} policy resulted in the mean confidence gain ranging from 2.6\% to 5.7\% and large effect sizes ($d > 0.30$). In contrast, the \textbf{Random} baseline yielded minimal or even negative mean gains in the UNSW domain, suggesting its unreliability despite some observed favourable p-values. In other conditions, we observed significant improvements over the \textit{Working Alone} baseline ($p < 0.0001$).

This analysis quantifies the value of collaboration between \systemname{} and simulated analysts. \textbf{Our findings demonstrate that collaboration with \systemname{} improves analyst performance compared to acting alone, even when adopting suggestions at random.} The adoption of confidence-based \systemname{} suggestions shows that \systemname{} contributes meaningful contextual information that enhances decision quality, irrespective of the analyst's confidence. 


Across both domains, \textbf{we observed a few instances (around 15 in total; 2 for HIKARI) where analysts initially made correct decisions but reversed them after accepting \systemname{} suggestion}. In HIKARI, one changed from TN to FP and the other from TP to TN. For UNSW, 10 changed from TN to FP, and three from TP to TN. These findings reveal that even highly accurate collaborative configurations can yield detrimental outcomes when analysts adopt suggestions uncritically (based on confidence). \textbf{Practically, this suggests that \systemname{} must provide explainability information to help analysts assess when not to follow or trust it}. 

\begin{figure}[h]
    \centering
    \includegraphics[width=0.9\linewidth]{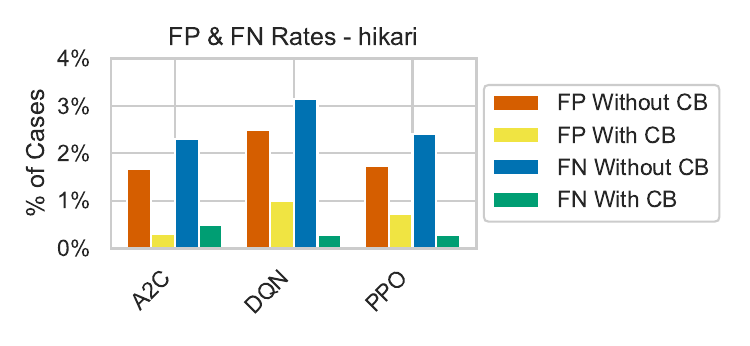}
    \caption{Improvement in FP and FN when simulated analysts engaged with \systemname{} (HIKARI).}
    \label{fig:fp_fn_rates_HIKARI}
\end{figure}

\subsection{\textbf{Reducing False Positives and Identifying More Attacks}} 

In the \textbf{HIKARI} domain (Figure~\ref{fig:fp_fn_rates_HIKARI}), analysts had many \textit{false positives} and \textit{false-negatives} (driven primarily by \textit{probing} events) when working alone. \textbf{\systemname{} helped analysts consistently cut false-positives, by 35 to 80\%, while also helping them correct occasional attack misclassification} (baseline F1 was already $\geq$\,0.96), especially for \textbf{probing} events, resulting in consistent gains of around 1.5\%. The teamwork resulted in all simulated analysts identifying all \textit{bruteforce}, \textit{bruteforce-xml} and almost all \textit{crypto miner} attacks that analysts initially misclassified. These results show that \systemname's suggested context cues helped analysts correctly classify real threats.

\begin{figure}[h]
    \centering
    \includegraphics[width=0.9\linewidth]{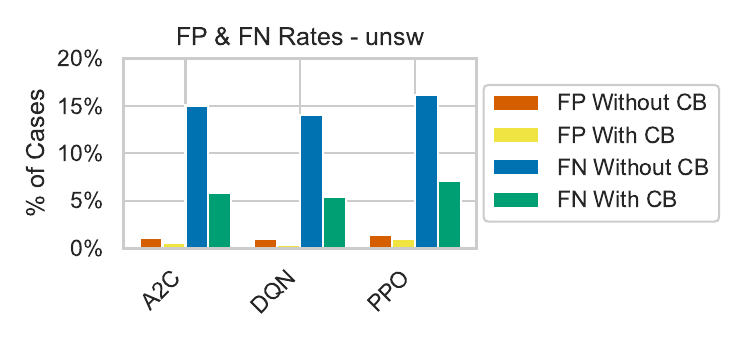}
    \caption{Improvement in FP and FN when simulated analysts engaged with \systemname{} (UNSW).}
    \label{fig:fp_fn_rates_UNSW}
\end{figure}

The \textbf{UNSW} domain (Figure~\ref{fig:fp_fn_rates_UNSW}) was substantially more complex for the simulated analysts than HIKARI: the trained RL analysts exhibited \textbf{poor performance on specific attacks} such as \emph{Exploits} and \emph{Backdoors}, with baseline $F_1 \approx 0.45$--$0.78$. Across all models, \textbf{\systemname{} helped analysts recover high‑impact attacks}: for \emph{Exploits}, \emph{Backdoors}, \emph{Generic}, and \emph{Analysis} flows, it \textbf{raised $F_{1}$ from 0.46–0.81 to 0.78–0.91}, and helped reduce the minimal FPs. 

Across domains, \systemname{} offered a mix of context types to assist, with none strongly correlating to any particular attack, underscoring its scenario‑specific guidance.

\begin{figure}[h]
    \centering
    \includegraphics[width=0.92\linewidth]{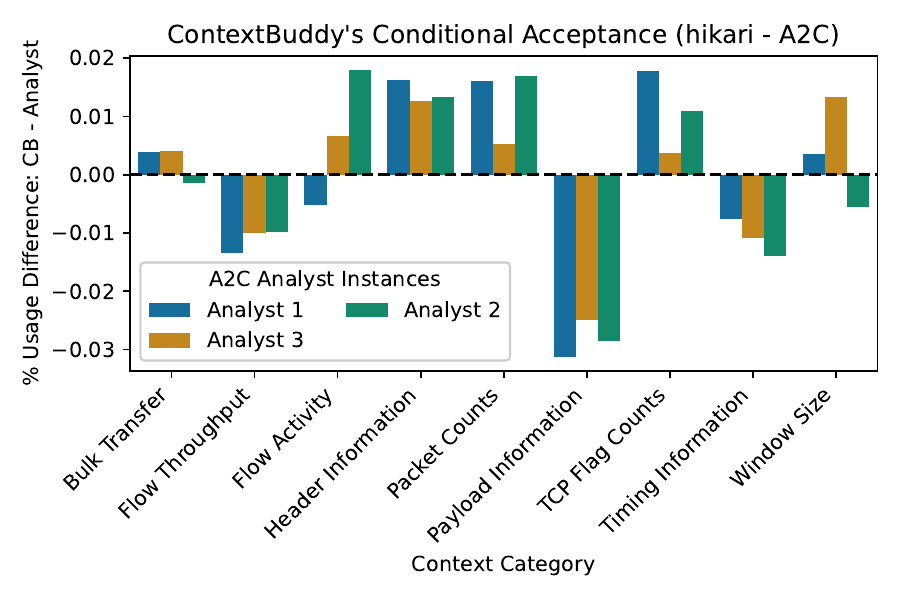}
    \caption{The figure shows how \systemname's suggestions were conditionally used by different analysts. Negative values show context more heavily used by analysts, while positive ones are additional ones suggested by \systemname{}.}
    \label{fig:context_diff_HIKARI_a2c}
\end{figure}

\subsection{\textbf{Context Usage Patterns}}
Finally, we examined how different simulated analysts and \systemname{} utilised context features during decision-making. The goal was to investigate whether \systemname{} merely replicated analyst behaviour or learned distinct usage patterns that influenced its recommendations. Figure~\ref{fig:context_diff_HIKARI_a2c}, shows an example from the HIKARI dataset for three A2C analysts working with \systemname{} and how it complements analyst decision-making by suggesting additional contextual features. Although \systemname's suggestions are not personalised to individual analysts, they often align with useful context, particularly in categories like \emph{Flow Activity}, which are more frequently adopted across all three analysts. Categories where \systemname{} consistently adds new, previously uncollected context, such as \emph{Header Information}, reveal how it can extend analyst reasoning and fill contextual gaps. \textbf{While \systemname's suggestions are generic, their value is realised through conditional interpretation and integration by different analysts, highlighting the synergistic potential of human-AI teaming in investigative workflows}. Note that different analysts may accept suggestions for different alerts, so Figure~\ref{fig:context_diff_HIKARI_a2c} shows a generalised trend.

\section{Evaluation (User Study)}

\begin{table*}[h]
\centering
\caption{Summary of Hypotheses}
\begin{tabular}{@{}llp{12cm}@{}}
\toprule
\textbf{Category} & \textbf{Hypothesis} & \textbf{Description} \\ \midrule
\multirow{2}{*}{Classification Accuracy} 
& H1a & C2 $\ge$ C1 as \systemname{} helps focus on relevant features, reducing irrelevant information. \\
& H1b & C3 $>$ C2 if participants effectively use extra features without becoming overwhelmed. \\

\multirow{2}{*}{Time Taken} 
& H2a & C2 $<$ C1 due to fewer, more relevant features speeding decisions. \\
& H2b & C2 $<$ C3; analysing extra info in C3 may increase decision time. \\

\multirow{2}{*}{Self-Confidence} 
& H3a & C1 $>$ C2 if participants view additional info as helpful. \\
& H3b & C3 $>$ C2 due to control over feature selection. \\

\multirow{2}{*}{Trust} 
& H4a & C1 $>$ C2 if explanations are perceived as relevant and useful. \\
& H4b & C3 $>$ C2, C1 as participants can validate and adjust \systemname's suggestions, increasing trust in it. \\

\multirow{2}{*}{Cognitive Load} 
& H5a & C1 $>$ C2 as analysing more features increases mental effort. \\
& H5b & C3 $>$ C2 due to effort managing interactivity and integrating extra features. \\
\bottomrule
\end{tabular}
\end{table*}

Having established promising results in the simulation experiments, this section discusses a follow-up user study designed for non-experts (experiments with real SOC analysts will be future work). We used the same version of \systemname{} as in the simulation study (\texttt{GAIL\_ALL}). We crafted two new research questions specific to the user study.

\subsection{Research Questions}
\begin{enumerate}
    \item How does \systemname{} impact the users' decision accuracy, efficiency, confidence, trust, and cognitive load?
    \item How does allowing users to request additional context beyond \systemname's suggestions affect these measures?
\end{enumerate}

\subsection{Independent Variables}
The user study assessed \systemname's ability to guide users towards relevant context information. Participants were given access to all features in the baseline setting. To assess \systemname's effectiveness, we restricted participants to only the most relevant features suggested by it. This was our primary objective. We expected improved accuracy and efficiency if \systemname{} was effective. We also allowed participants to toggle between all features and \systemname-selected ones. However, this was a secondary goal, as this feature would benefit experts in intrusion detection.

\begin{enumerate}
    \item \textbf{Analyst-only (C1)}: In this baseline, participants see all features with additional information (see Sec~\ref{sec:user-study-task}). 
    \item \textbf{One-time Assistance (C2)}: The participants use only feature subsets \systemname{} provides them. 
    \item \textbf{Iterative Assistance (C3)}: \systemname{} suggests features, but participants can modify these suggestions. 
\end{enumerate}

\subsection{Dependent Variables}
\begin{enumerate}
    \item \textbf{Classification Accuracy (objective)}: percentage of correct classifications (against the ground truth).
    \item \textbf{Time Taken (objective)}: The system automatically tracks participants' time to classify each alert.
    \item \textbf{Self-Confidence (subjective)}: Participants' confidence in their classifications; measured on a 0\% to 100\% scale.
    \item \textbf{Trust (subjective)}: Participants' trust \textit{in the AI generated explanations}; measured using the scale by \cite{Jian2000-nm}\footnote{See Appendix E for list of trust related questions}.
    \item \textbf{Cognitive Load (subjective)}: Participants' perceived cognitive load; assessed using Klepsch's instrument~\cite{Klepsch2017-fj}\footnote{See Appendix E for list of cognitive load related questions}.
\end{enumerate}

\subsection{Tasks} \label{sec:user-study-task}
Approved by the relevant Human Ethics Committee, the user studies included two pilot studies to ensure participants understood the tasks and the dashboard interface (see Figs 12 - 19 in the appendix). Participants classified alerts into attack types using a dashboard that presented them with alerts. Participants reviewed them sequentially. Clicking on an alert revealed \textit{all features} (baseline) or a subset selected by \systemname{} depending on the condition. 

\systemname{} provided explanations for selected features, computed using SHAP-based XAI\footnote{https://shap.readthedocs.io/en/latest/generated/shap.TreeExplainer.html}. We computed explanations for all context combinations and visualised \systemname's selected feature importance, showing each feature’s positive/negative contributions and per-class summaries.

Given their non-expert status, participants needed support to build a mental model for each alert's most likely class. We provided each feature's historical feature statistics (mean, median, mode), computed solely from historical data (without AI input). These statistics helped participants compare the current alert's features with historical patterns. This source of information allowed participants to evaluate \systemname's explanations critically. The dual-source approach (feature statistics alongside SHAP-based explanations) encouraged active engagement with the data and prevented users from passively accepting the \systemname's evidence.

\subsection{Procedure} We had two phases: familiarisation and testing. 

\noindent\textbf{Familiarisation Phase:} Participants began with a 30-minute familiarisation phase, during which we introduced them to the task through a practice instance to help them understand the system interface, including how to analyse feature visualisations and interpret AI-generated explanations. 

\noindent\textbf{Testing Phase:} Participants then proceeded to the testing phase, which consisted of three conditions. Participants classified 4 alerts in each condition, selected to represent different classification challenges: 1) 1 True Positive (TP) with high confidence (easy classification); 2) 1 True Negative (TN) with high confidence; 3) 1 False Positive (FP) with low confidence (challenging classification due to AI confusion between classes); and 4) 1 False Negative (FN) with low confidence. We randomly selected 12 alerts that matched the above classification outcomes from the instances not used in the simulation study. 

We selected the alerts for each condition to minimise the pairwise distance between corresponding instances (e.g., the TP alerts across all conditions are very similar) because we could not use the same instances across the conditions. This approach ensures that differences in participant performance across conditions are due to condition-specific factors rather than variations in the alerts themselves. We divided the 12 alerts into 3 datasets, each with 4 alerts\footnote{See Table VII in the appendix for predicted probabilities}. We randomised the order of the 4-question datasets (for C1 and C2; see Sec~\ref{sec:user-study-design-considerations}) and the order of the 4 alerts within each condition to reduce the ordering effect. Participants spent approximately 20 minutes per condition. They completed the trust and cognitive load questionnaires at the end of each condition and took short breaks between conditions to help maintain focus.

\subsubsection{Design Considerations}
\label{sec:user-study-design-considerations}

Conditions 1 and 2 were counterbalanced to reduce order effects, while C3 was always placed last. This arrangement prevented participants from knowing about the feature selection option in advance, ensuring unbiased evaluations in the earlier conditions. However, placing C3 last could impact performance due to fatigue, changes in strategy because of the new option to select additional features, or lack of domain expertise (we had non-experts, i.e. non-SOC analysts).

\section{Results (User Study)}
In this section, we discuss the results of our user study.

\small
\begin{table}[h]
\centering
\begin{tabular}{|p{1.2cm} c c c c|}
\hline
\textbf{Condition} & \textbf{CB} & \textbf{Participant} & \textbf{One-time} & \textbf{Iterative} \\
\hline
Accuracy & 75.0\% & 65.4\% & 86.5\% & 65.4\% \\
\hline
Time (sec per alert) & - & 187 $\pm$ 120 & 142 $\pm$ 108 & 183 $\pm$ 116 \\
\hline
Confidence & 82.9 $\pm$ 16.1 & 79.9 $\pm$ 18.0 & 84.6 $\pm$ 12.7 & 79.6 $\pm$ 17.3 \\
\hline
Trust  & - & 5.23 $\pm$ 0.86 & 4.95 $\pm$ 0.76 & 4.94 $\pm$ 1.02 \\
\hline
Cog Load & -  & 3.53 $\pm$ 0.74 & 3.36 $\pm$ 0.64 & 3.88 $\pm$ 0.75 \\
\hline
\hline
\multicolumn{5}{|c|}{\textbf{Accuracy by Classification Outcome (\%)}}\\
\hline
TP & 100 & 84.6 & 100 & 100 \\
TN & 66.7 & 92.3 & 100 & 100 \\
FP & 33.3 & 0 & 53.8 & 0 \\
FN & 100 & 84.6 & 92.3 & 61.5 \\
\hline
\hline
\multicolumn{5}{|c|}{\textbf{Overall Information Usage (Max 5)}} \\
\hline
Features  & -  & 4.31 $\pm$ 0.85 & 4.38 $\pm$ 0.79 & 3.81 $\pm$ 1.10 \\
Explanations  & -  & 4.06 $\pm$ 0.96 & 4.19 $\pm$ 0.97 & 4.12 $\pm$ 1.06 \\
Knowledge  & -  & 1.94 $\pm$ 0.89 & 1.96 $\pm$ 1.05 & 2.06 $\pm$ 1.06 \\
\hline

\end{tabular}
\caption{User study results. One-time and Iterative conditions are where participants teamed with ~\systemname to classify intrusion events. CB = \systemname~only.}
\label{tab:results_summary}
\end{table}

\normalsize

Table~\ref{tab:results_summary} shows the main results\footnote{
7 participants did $C1 \rightarrow C2$; others $C2  \rightarrow C1$; 7 classified 4 alerts from dataset 1 then 4 from dataset 2; others did the reverse. C3 order and alerts were same.}. \textbf{Each participant session involved 12 different alerts and lasted for 90 - 120 minutes}. As before, we used a Wilcoxon signed-rank test with Bonferroni correction and reported only noteworthy cases. \textbf{We  performed a post-hoc power analysis, achieving 0.77 - 0.81 for $\alpha = 0.05$ and moderate effects (we achieved large effects)}. \textbf{Demographics:} Of 13 participants, most (10) had technical knowledge of computer networks, while 3 did not. Only one participant had prior experience with intrusion detection systems, reporting 4 years of use; the remaining 12 had not used IDS. Regarding proficiency in network and packet analysis tools, 2 participants considered themselves very knowledgeable, 2 moderately knowledgeable, 3 slightly knowledgeable, and 6 not knowledgeable at all.

\subsubsection{\textbf{Classification Accuracy}}
Participants in the One-time condition achieved the highest accuracy at 86.5\%, compared to 65.4\% in the Analyst and 65.4\% in the Iterative conditions. Analyst and One-time comparison showed a significant difference: $p=0.008~ (Z = 2.994, r = 0.831)~pwr=0.81$. Similarly, the Iterative and One-time comparisons showed a significant difference: $p=0.018~ (Z = 2.747, r = 0.762)~pwr=0.77$ (unadjusted $p=0.006$). \textbf{The significant increase from Analyst to One-time conditions supports \textbf{H1a}, indicating that \systemname's guidance enhances accuracy by helping users focus on relevant features}. Contrary to \textbf{H1b}, participants' accuracy in the Iterative condition decreased. As users saw C3 last, this placement impacted performance. 

\textbf{Regarding TN and TP, participants in the two assisted conditions achieved 100\% accuracy}, outperforming the analyst condition (TN: 92.3\%, TP: 84.6\%). For FN, accuracy was best in One-time condition (92.3\%). In FP, participants in the One-time condition improved to 53.8\% accuracy, while other conditions remained at 0\%. 

\subsubsection{Completion Time} \textbf{Participants in the One-time condition completed tasks faster} (142~ $\pm$ ~108 seconds per alert), compared to 187~$\pm$~120s in the Analyst and 183~$\pm$~116s in the Iterative conditions. Analyst and One-time comparison resulted in marginal significance: $p = 0.014~(Z = 2.831, r = 0.785)$ (unadjusted $p=0.005$). This supports \textbf{H2a}, indicating that \textbf{focusing on relevant features reduces decision-making time}. Since toggling between all and relevant features notably increases completion times ($142 \rightarrow 183$) in the Iterative condition, results partially support \textbf{H2b} ($p=0.09)$.

\subsubsection{Qualitative Feedback from Participants}
Post-study discussions revealed that participants perceived the Iterative condition as offering \textit{two versions} of \systemname{}, one showing all features and another showing only the most relevant ones. In TN and TP cases, both versions aligned in confidence and evidence, making decisions straightforward; all participants were 100\% accurate in these scenarios. In contrast, FP and FN cases involved different or uncertain evidence between the \textit{two versions}. This uncertainty prompted participants to revisit feature visualisations and explanations, and final decisions varied; some followed \systemname{}, and others relied on their own interpretation of all features. Although this was partially due to lack of expertise, to reduce inappropriate adoption based on marginal differences in confidence or AI evidence strength, future versions of \systemname{} could explore frequency-based confidence visualisations~\cite{cao2024designing}, which have been shown to help users better trust calibration.

\section{Discussion}
\label{sec:discussion}

Our study investigated the potential of \systemname{}, an AI designed to function in a human-AI team and support SOC analysts by suggesting relevant contextual information during alert investigations and, in this paper, specifically for validating intrusion events. There were a few core requirements to demonstrate the success of \systemname{}: investigation data, learning method to be used by \systemname{} to learn analysts contextual information use, and reasoning and interaction mechanisms to facilitate the transfer of contextual information identified by \systemname{} for specific events to analysts.

We first trained various RL agents (A2C, PPO, DQN) to generate investigation traces, forming the basis of training \systemname{}. While \systemname{} could be trained on any appropriate method, we tried various IL methods (BC, AIRL, GAIL). \textbf{The results demonstrate the effectiveness of IL, particularly GAIL, in modelling diverse analyst context-seeking behaviours}. Importantly, we observed significant performance gains when utilising multi-source GAIL (when GAIL provided traces from analysts trained on all three models), \textbf{suggesting diversity of analyst behaviours is essential for better generalisation}. This aligns with previous literature highlighting the advantages of adversarial imitation in capturing nuanced behavioural patterns in dynamic decision-making environments~\cite{Ho2016-uc,Fu2017-rc}.

Our simulation study, where simulated analysts had access to \systemname{}, demonstrates that \textbf{\systemname{} can meaningfully improve analyst performance across several collaboration strategies we tested}. For example, when analysts adopted \systemname's contextual suggestions when uncertain, they performed better than working alone. Notably, \textbf{\systemname{} helped analysts reduce both false positives (by 35–80\% in HIKARI dataset) and false negatives (raising F$_1$ scores from as low as 0.46 to as high as 0.91 in UNSW dataset)}. However, \textbf{collaboration is not without risk}. Analysts accepted \systemname's suggestion in a few cases and made a worse decision. This highlights the importance of calibrated adoption and suggests that AI assistants like \textbf{\systemname{} should be paired with actionable explanations, as we did in the user study, to help analysts decide when to trust or override suggestions}. Well-calibrated, human-in-the-loop AI support can deliver measurable improvements without sacrificing control.

Moreover, \systemname{} \textbf{demonstrated substantial practical utility in our user study even for non-expert analysts, where participants achieved a 21.1\% increase in classification accuracy and a 24\% reduction in alert validation time}, especially under the one-time assistance mode. The observed performance degradation under the iterative mode suggests that \textbf{iterative interactions} may require higher analyst expertise, and for \textbf{\systemname{} to cater for distinct decision-maker styles~\cite{lu2024mix} and support better integration of confidence in decision-making}, e.g via frequency-based presentations~\cite{cao2024designing}.

Overall, the paper provides a very promising \textbf{demonstration of how an AI assistant like \systemname{} can augment analyst decision-making through context-aware support}, highlighting the value of IL for modelling human investigation behaviours and the importance of adaptive collaboration strategies in operationalising human-AI teaming in SOCs.

\section{Limitations and Future Work}

While these results highlight the clear value of \systemname{}, limitations must be acknowledged. We used simulated analysts; naturally, we need real-world SOC data. Similarly, the user study involved non-experts, limiting generalisability to experienced SOC analysts. Future work will address these gaps through studies with professional analysts. Notably, \systemname's modular design supports extensibility beyond IL. Given the growing deployment of LLMs~\cite{Freitas2024-sx} future work will explore it as option for replacing IL. However, as noted earlier, the performance reported earlier (F1 less than 0.74 on UNSW) and issues with potential of LLMs to mislead users with convincing explanations~\cite{kim2025fostering} demands caution. Currently, we would need to retrain \systemname{} models periodically; combining imitation with real-time human feedback or active learning, as seen in systems like AlertPro~\cite{Wang2024-zz}, may enhance performance in dynamic SOC environments.

\section{Conclusion}
\label{sec:conclusion}

Our work contributes a methodological foundation for systematically enhancing SOC operations through intelligent human-AI teaming. By offering tailored contextual information guidance, \systemname~ promises to address the critical issue of lack of context in SOC, improve accuracy, reduce false positives and false negatives, and reduce validation time. These can help reduce alert fatigue as well. We anticipate that \systemname-like assistants will be a foundational platform for future advancements in AI-assisted investigations, paving the way for collaborative and efficient cybersecurity operations.

\bibliographystyle{ieeetr}
\bibliography{references}

\appendix
\section{Appendix}

\section*{Appendix: Figures and Tables Summary}

\begin{itemize}
    \item \textbf{Figure 10: Class Distribution of Selected Alerts (HIKARI Dataset)}  
    \begin{itemize}
        \item \textit{Purpose:} Shows percentage of each class in historical vs. new HIKARI alerts.
    \end{itemize}

    \item \textbf{Figure 11: Class Distribution of Selected Alerts (UNSW-NB15 Dataset)}  
    \begin{itemize}
        \item \textit{Purpose:} Displays class percentages for historical and new alerts in UNSW-NB15.
    \end{itemize}

    \item \textbf{Figure 12: High-level Overview of Participant Task}  
    \begin{itemize}
        \item \textit{Purpose:} Summarises user study procedure for classifying alerts.
    \end{itemize}

    \item \textbf{Figure 13: Table Display of Alerts}  
    \begin{itemize}
        \item \textit{Purpose:} Shows how participants view raw alerts.
    \end{itemize}

    \item \textbf{Figure 14: Example Feature Visualisation (mean, median, mode)}  
    \begin{itemize}
        \item \textit{Purpose:} Depicts summary stats for alert feature inspection.
    \end{itemize}

    \item \textbf{Figure 15: Explanation of feature visualisation}  
    \begin{itemize}
        \item \textit{Purpose:} Provides instructions to participants on how to interpret feature statistics (mean, median, mode) displayed for each alert feature. Supports consistent understanding of visualised data during classification.
    \end{itemize}

    \item \textbf{Figure 16: Iterative Panel}  
    \begin{itemize}
        \item \textit{Purpose:} Displays the interface component used in the Iterative (C3) condition, allowing participants to filter and explore both ContextBuddy-suggested and self-selected features for investigation.
    \end{itemize}

    \item \textbf{Figure 17: AI-generated explanations displaying feature importance values for different classes}  
    \begin{itemize}
        \item \textit{Purpose:} Visualises SHAP-based explanations that quantify each feature’s contribution (positive or negative) toward a class prediction, aiding participant interpretation of AI guidance.
    \end{itemize}

    \item \textbf{Figure 18: Discussion of how to interpret explainability information}  
    \begin{itemize}
        \item \textit{Purpose:} Offers explanatory guidance on how users should understand and critically assess AI explanations when making classification decisions.
    \end{itemize}

    \item \textbf{Figure 19: Final classification decision panel}  
    \begin{itemize}
        \item \textit{Purpose:} Shows the submission screen where participants choose their final alert classification, rate confidence, and indicate reliance on explanations or visualisations.
    \end{itemize}

    \item \textbf{Figure 20: User accuracies}  
    \begin{itemize}
        \item \textit{Purpose:} Reports the classification accuracy achieved by participants across different study conditions, reflecting the effectiveness of ContextBuddy support.
    \end{itemize}

    \item \textbf{Figure 21: Completion Times}  
    \begin{itemize}
        \item \textit{Purpose:} Compares average task completion times across three experimental conditions (Analyst-only, One-time ContextBuddy, Iterative ContextBuddy), showing how different forms of AI support affect efficiency.
    \end{itemize}

    \item \textbf{Figure 22: User Confidence}  
    \begin{itemize}
        \item \textit{Purpose:} Visualises self-reported confidence levels for participant decisions in each condition, indicating the perceived decisional certainty with or without AI assistance.
    \end{itemize}

    \item \textbf{Figure 23: Aggregated Trust}  
    \begin{itemize}
        \item \textit{Purpose:} Summarises trust ratings from participants regarding the AI explanations and its behaviour, highlighting differences in trust across assistance modes.
    \end{itemize}

    \item \textbf{Figure 24: Aggregated Cognitive Load}  
    \begin{itemize}
        \item \textit{Purpose:} Plots perceived cognitive load experienced by users in each condition, providing insights into mental effort required when working with or without ContextBuddy.
    \end{itemize}

    \item \textbf{Figure 25: Simulated Analyst Training Process}  
    \begin{itemize}
        \item \textit{Purpose:} Outlines the sequential steps in training simulated analysts to classify alerts, including context requests and classification using XGBoost, forming the foundation for training ContextBuddy.
    \end{itemize}

    \item \textbf{Figure 26: ContextBuddy (AIRL Assistant) Training Process}  
    \begin{itemize}
        \item \textit{Purpose:} Visualises how ContextBuddy learns to imitate analyst decision-making using Adversarial Inverse Reinforcement Learning (AIRL), including data collection, training, and generalisation to suggest context features for new alerts.
    \end{itemize}

    \item \textbf{Figure 27: Testing the Simulated Analyst-ContextBuddy Dyad}  
    \begin{itemize}
        \item \textit{Purpose:} Depicts the workflow of collaborative alert classification, where ContextBuddy assists a simulated analyst by suggesting context during SOC-like alert investigations.
    \end{itemize}

    \item \textbf{Figure 28: Breakdown of Performance by Individual Analysts (HIKARI, Threshold 0.6)}  
    \begin{itemize}
        \item \textit{Purpose:} Shows weighted F1 scores of three RL analysts working with ContextBuddy, adopting AI suggestions only when their confidence is below 0.6 on the HIKARI dataset.
    \end{itemize}

    \item \textbf{Figure 29: Breakdown of Performance by Individual Analysts (HIKARI, Threshold 0.9)}  
    \begin{itemize}
        \item \textit{Purpose:} Presents F1 performance across three RL analysts with a higher confidence threshold (0.9) for accepting ContextBuddy suggestions, showing greater performance shifts.
    \end{itemize}

    \item \textbf{Figure 30: Breakdown of Performance by Individual Analysts (UNSW, Threshold 0.6)}  
    \begin{itemize}
        \item \textit{Purpose:} Illustrates how simulated analysts performed under low-confidence threshold-based adoption of AI context advice in the UNSW dataset setting.
    \end{itemize}

    \item \textbf{Figure 31: Breakdown of Performance by Individual Analysts (UNSW, Threshold 0.9)}  
    \begin{itemize}
        \item \textit{Purpose:} Compares the weighted F1 scores of three RL analysts under the 0.9 confidence threshold for adopting ContextBuddy suggestions, in the UNSW dataset.
    \end{itemize}

    \item \textbf{Figure 32: Simulated Analysts’ Confidence Gain (HIKARI)}  
    \begin{itemize}
        \item \textit{Purpose:} Plots average prediction confidence across suggestion acceptance strategies (e.g., random, always accept, threshold-based) in the HIKARI domain.
    \end{itemize}

    \item \textbf{Figure 33: Simulated Analysts’ Confidence Gain (UNSW)}  
    \begin{itemize}
        \item \textit{Purpose:} Shows confidence changes for different analyst-AI collaboration strategies in the UNSW domain, highlighting effectiveness of ContextBuddy support.
    \end{itemize}
\end{itemize}

\begin{itemize}
    \item \textbf{Table IV: HIKARI-2021 Feature Groupings}  
    \begin{itemize}
        \item \textit{Purpose:} Lists context categories for the HIKARI-2021 dataset, detailing which features belong to each and their diagnostic relevance for alert classification (e.g., Initial Features, Packet Counts, TCP Flags, etc.).
    \end{itemize}

    \item \textbf{Table V: UNSW-NB15 Feature Groupings}  
    \begin{itemize}
        \item \textit{Purpose:} Presents feature categories and corresponding attributes from the UNSW-NB15 dataset, explaining what each category captures and its importance for intrusion detection (e.g., Timing Information, State Information).
    \end{itemize}

    \item \textbf{Table VI: Example Alert Investigation Trajectory}  
    \begin{itemize}
        \item \textit{Purpose:} Describes a step-by-step example of how a simulated analyst sequentially requests context features to refine classification, showing time steps, feature sets, and model confidence progression.
    \end{itemize}

    \item \textbf{Table VII: Alerts Used in the User Study with Predicted Probabilities}  
    \begin{itemize}
        \item \textit{Purpose:} Summarises the alerts shown to participants during the user study, including alert IDs, predicted probabilities from the classifier, predicted and true classes, and final classification outcome (TP, TN, FP, FN).
    \end{itemize}

    \item \textbf{Appendix E: Trust and Cognitive Load Questionnaire Items}  
    \begin{itemize}
        \item \textit{Purpose:} Lists items used to measure participants’ subjective trust in AI explanations and perceived cognitive effort. Based on validated scales (e.g., Jian et al. and Klepsch et al.).
    \end{itemize}
\end{itemize}

\begin{table*}[h]
\centering
\small
\setlength{\tabcolsep}{4pt}
\begin{tabular}{@{}l p{0.35\textwidth} p{0.45\textwidth}@{}}
\toprule
\textbf{Category} & \textbf{Features (HIKARI‑2021 field names)} & \textbf{What the category represents / why it matters} \\ \midrule

Initial Features &
\texttt{originp}, \texttt{responp}, \texttt{flow\_duration},
\texttt{fwd\_pkts\_tot}, \texttt{bwd\_pkts\_tot}, \texttt{down\_up\_ratio},
\texttt{flow\_SYN\_flag\_count}, \texttt{flow\_RST\_flag\_count}, \texttt{flow\_ACK\_flag\_count},
\texttt{active.avg}, \texttt{idle.avg} &
Core flow metadata supplied to RL agents as their starting observation: ports, duration, packet counts, traffic asymmetry, key TCP flag totals, and mean active/idle periods. \\[0.5em]

Packet Counts &
\texttt{fwd\_data\_pkts\_tot}, \texttt{bwd\_data\_pkts\_tot}, \texttt{fwd\_pkts\_per\_sec}, \texttt{bwd\_pkts\_per\_sec}, \texttt{flow\_pkts\_per\_sec} &
Directional packet volumes and per‑second rates. \\[0.5em]

Header Information &
\texttt{fwd/bwd\_header\_size\_[tot|min|max]} &
Sizes of protocol headers exchanged.  \\[0.5em]

TCP Flag Counts &
\texttt{flow\_FIN\_flag\_count},
\texttt{fwd/bwd\_PSH\_flag\_count},
\texttt{fwd/bwd\_URG\_flag\_count},
\texttt{flow\_CWR\_flag\_count}, \texttt{flow\_ECE\_flag\_count} &
Control‑plane flag statistics.  \\[0.5em]

Payload Information &
\texttt{fwd/bwd/flow\_pkts\_payload.[min|max|tot|avg|std]} &
Distribution of user‑data sizes per direction and overall.  \\[0.5em]

Timing Information &
\texttt{fwd/bwd/flow\_iat.[min|max|tot|avg|std]} &
Inter‑arrival‑time statistics.  \\[0.5em]

Flow Throughput &
\texttt{payload\_bytes\_per\_second}, \texttt{fwd/bwd\_subflow\_pkts}, \texttt{fwd/bwd\_subflow\_bytes} &
Instantaneous byte rate and logical sub‑flow segmentation.  \\[0.5em]

Bulk Transfer Properties &
\texttt{fwd/bwd\_bulk\_bytes},
\texttt{fwd/bwd\_bulk\_packets},
\texttt{fwd/bwd\_bulk\_rate} &
Counters derived from consecutive large segments.  \\[0.5em]

Flow Activity &
\texttt{active.[min|max|tot|std]},
\texttt{idle.[min|max|tot|std]} &
Durations of active transmission versus idle pauses. \\[0.5em]

Window Size Information &
\texttt{fwd\_init\_window\_size}, \texttt{bwd\_init\_window\_size}, \texttt{fwd\_last\_window\_size} &
Advertised TCP window sizes at key points.  \\ \bottomrule
\end{tabular}
\caption{HIKARI‑2021 feature groupings, their constituent attributes, and their diagnostic relevance.}
\label{tab:hikari_feature_groups}
\end{table*}

\begin{figure}[h]
    \centering
    \includegraphics[width=\columnwidth]{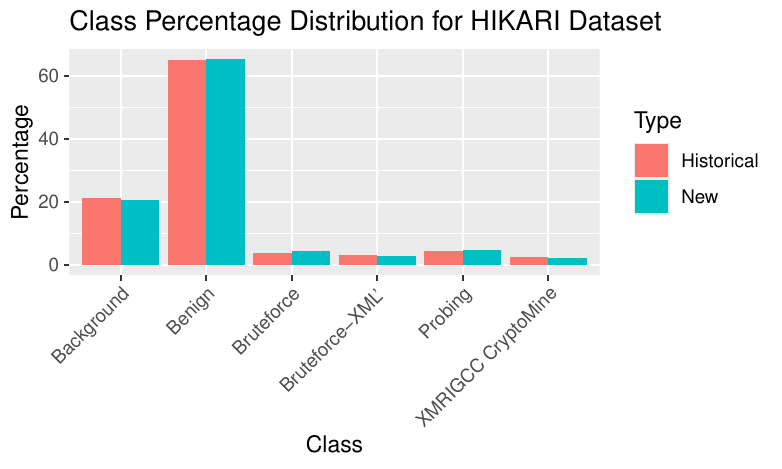}  
    \caption{Class Distribution of selected alerts matching the distribution in the original file.}
    \label{fig:class-distrubtion-hikari}
\end{figure}

\subsection{HIKARI-2021 Dataset Information}
\label{sec:hikari_dataset_info}

The following lists the context categories generated for the Hikari dataset. We show the features provided to the simulated analysts to start the investigation and the features that fall under each context category in Tab~\ref{tab:hikari_feature_groups}. We show the class distribution of the selected alerts in Fig~\ref{fig:class-distrubtion-hikari}.

\begin{table*}[h]
\centering
\small
\setlength{\tabcolsep}{4pt}
\begin{tabular}{@{}l p{0.35\textwidth} p{0.45\textwidth}@{}}
\toprule
\textbf{Category} & \textbf{Features (UNSW‑NB15 field names)} & \textbf{What the category represents} \\ \midrule
Initial Features &
\texttt{sbytes}, \texttt{dbytes}, \texttt{Spkts}, \texttt{Dpkts}, \texttt{Sload}, \texttt{Dload} &
These are the initial features provided to the agent to start the investigation. Then the RL agents request the following additional contexts to classify each intrusion event \\[0.3em]

Connection Dynamics &
\texttt{dur}, \texttt{Stime}, \texttt{Ltime}, \texttt{ct\_state\_ttl}, \texttt{sport}, \texttt{dsport} \\[0.3em]

PacketCounts & \texttt{smeansz}, \texttt{dmeansz}, \texttt{sloss}, \texttt{dloss} &
Average packet sizes sent/received and packet‑loss counters. \\[0.3em]

NetworkServiceUsage &
\texttt{service\_dhcp}, \texttt{service\_dns}, \texttt{service\_ftp}, \texttt{service\_ftp-data}, \texttt{service\_http}, \texttt{service\_irc}, \texttt{service\_pop3}, \texttt{service\_radius}, \texttt{service\_smtp}, \texttt{service\_snmp}, \texttt{service\_ssh}, \texttt{service\_ssl} &
One‑hot flags for well‑known application services. \\[0.3em]

WindowSize & \texttt{swin}, \texttt{dwin} &
TCP window sizes are negotiated by each endpoint. \\[0.3em]

ProtocolSpecificFeatures1 &
71 fine‑grained protocol indicators beginning with \texttt{proto\_3pc}–\texttt{proto\_leaf-2} &
Rare or legacy L3/L4 protocols. Splitting into two columns keeps the sparse vectors manageable for learners. \\[0.3em]

ProtocolSpecificFeatures2 &
73 indicators from \texttt{proto\_merit-inp}–\texttt{proto\_zero} (incl.\ \texttt{proto\_tcp}, \texttt{proto\_udp}, \texttt{proto\_icmp}) &
Common plus remaining protocols. \\[0.3em]

TimingInformation &
\texttt{sttl}, \texttt{dttl}, \texttt{Sjit}, \texttt{Djit}, \texttt{Sintpkt}, \texttt{Dintpkt}, \texttt{tcprtt}, \texttt{synack}, \texttt{ackdat} &
Time‑to‑live, jitters, inter‑packet gaps, TCP RTT and handshake latencies. \\[0.3em]

Relation &
\texttt{ct\_srv\_src}, \texttt{ct\_srv\_dst}, \texttt{ct\_dst\_ltm}, \texttt{ct\_src\_ltm}, \texttt{ct\_src\_dport\_ltm}, \texttt{ct\_dst\_sport\_ltm}, \texttt{ct\_dst\_src\_ltm} &
Temporal counts of connections between hosts/ports within a sliding “last‑time‑minute” window. \\[0.3em]

FlowBehaviour &
\texttt{ct\_ftp\_cmd}, \texttt{trans\_depth}, \texttt{res\_bdy\_len}, \texttt{is\_ftp\_login}, \texttt{is\_sm\_ips\_ports}, \texttt{ct\_flw\_http\_mthd}, \texttt{stcpb}, \texttt{dtcpb} &
Application‑layer semantics (FTP commands, HTTP method diversity), transaction depth, and TCP sequence numbers. \\[0.3em]

StateInformation &
\texttt{state\_ACC}, \texttt{state\_CLO}, \texttt{state\_CON}, \texttt{state\_ECO}, \texttt{state\_ECR}, \texttt{state\_FIN}, \texttt{state\_INT}, \texttt{state\_MAS}, \texttt{state\_PAR}, \texttt{state\_REQ}, \texttt{state\_RST}, \texttt{state\_TST}, \texttt{state\_TXD}, \texttt{state\_URH}, \texttt{state\_URN}, \texttt{state\_no} &
Connection states assigned by the Bro/Zeek parser. \\ \bottomrule
\end{tabular}
\caption{UNSW‑NB15 feature groupings, their constituent attributes, and their diagnostic relevance for intrusion detection.}
\label{tab:unsw_feature_groups}
\end{table*}

\begin{figure}[h]
    \centering
    \includegraphics[width=\columnwidth]{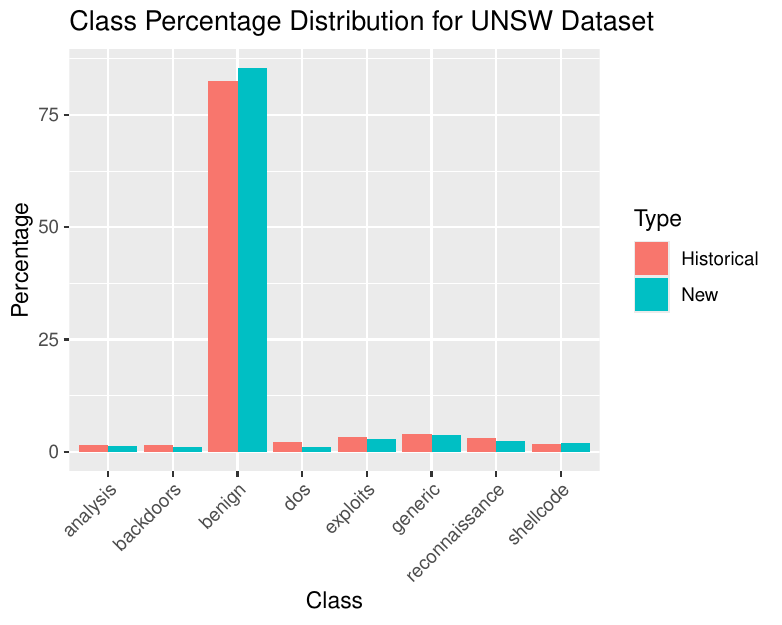}
    \caption{Class distribution of selected alerts for the UNSW dataset matching the distribution in the original file.}
    \label{fig:class-distribution-unsw}
\end{figure}

\subsection{UNSWNB-15 Dataset}
\label{sec:unsw_dataset_info}
The following lists the context categories generated for the UNSWNB-15 dataset. We show the initial features provided to the simulated analysts to start the investigation and the features that fall under each context category in Tab~\ref{tab:unsw_feature_groups}. We also show the class distribution of the selected alerts in Fig~\ref{fig:class-distribution-unsw}.

\subsection{Example Investigation Steps}
\label{sec:example_investigation_steps}
An example of an alert investigation performed by an analyst is shown in Tab~\ref{tab:example_investigation}.

\small
\begin{table*}[ht]
\caption{An example of alert investigation. The analysts start with an initial feature set provided by the SOC agent at the time, $t=0$. Then it requests packet counts ($t=1$) followed by payload information ($t=2$). Based on this additional contextual information, the analyst correctly classifies this alert to be \textit{benign} with 90\% confidence at $t=2$, ending the investigation. The probabilities are computed using an XGBoost classifier.}
    \label{tab:example_investigation}   
    \centering
    \begin{tabular}{|p{1.5cm}|p{4.5cm}|p{4.5cm}|p{4.5cm}|}
        \toprule
        \textbf{time} & $t=0$ & $t=1$ & $t=2$ \\
        \midrule
        \textbf{action} & get packet counts (pc) & get payload info (pl) & classify event \\
        \midrule
        
        \textbf{state} & $s_0$ & $s_1$ & $s_2$ \\
        \midrule
         \textbf{initial features} & $originp = 50967, responp = 53.0, flow\_duration = 0.02591, fwd\_pkts\_tot = 2.0, \ldots$ &

         $originp = 50967, responp = 53.0, flow\_duration = 0.02591, fwd\_pkts\_tot = 2.0, \ldots$   &

         $originp = 50967, responp = 53.0, flow\_duration = 0.02591, fwd\_pkts\_tot = 2.0, \ldots$   \\
        \midrule
         \textbf{context features} & none &
         $Packet Counts: \{fwd\_data\_pkts\_tot = 2.0 , bwd\_data\_pkts\_tot = 2.0 , fwd\_pkts\_per\_sec = 77.19, \ldots$\} &  
         $Packet Counts: \{fwd\_data\_pkts\_tot = 2.0 , bwd\_data\_pkts\_tot = 2.0 , fwd\_pkts\_per\_sec = 77.19, \ldots$\} AND 
         $Payload Information: \{fwd\_pkts\_payload.min = 36.0, fwd\_pkts\_payload.max = 36.0, \ldots \}$  
         \\
        \midrule
        \textbf{context encoding} & $pc=0; pl=0; \ldots$ & $pc=1; pl=0; \ldots$ & $pc=1; pl=1; \ldots$ \\
        \midrule
        \textbf{ground truth prob} & n/a & $prob_{dt}(benign | s_1)=0.89$ & $prob_{dt}(benign | s_2)=1.0$\\
         
        \bottomrule

    \end{tabular}
    
\end{table*}

\section{User Study Related}

\begin{figure*}[h]
    \centering
    \includegraphics[width=1.0\linewidth]{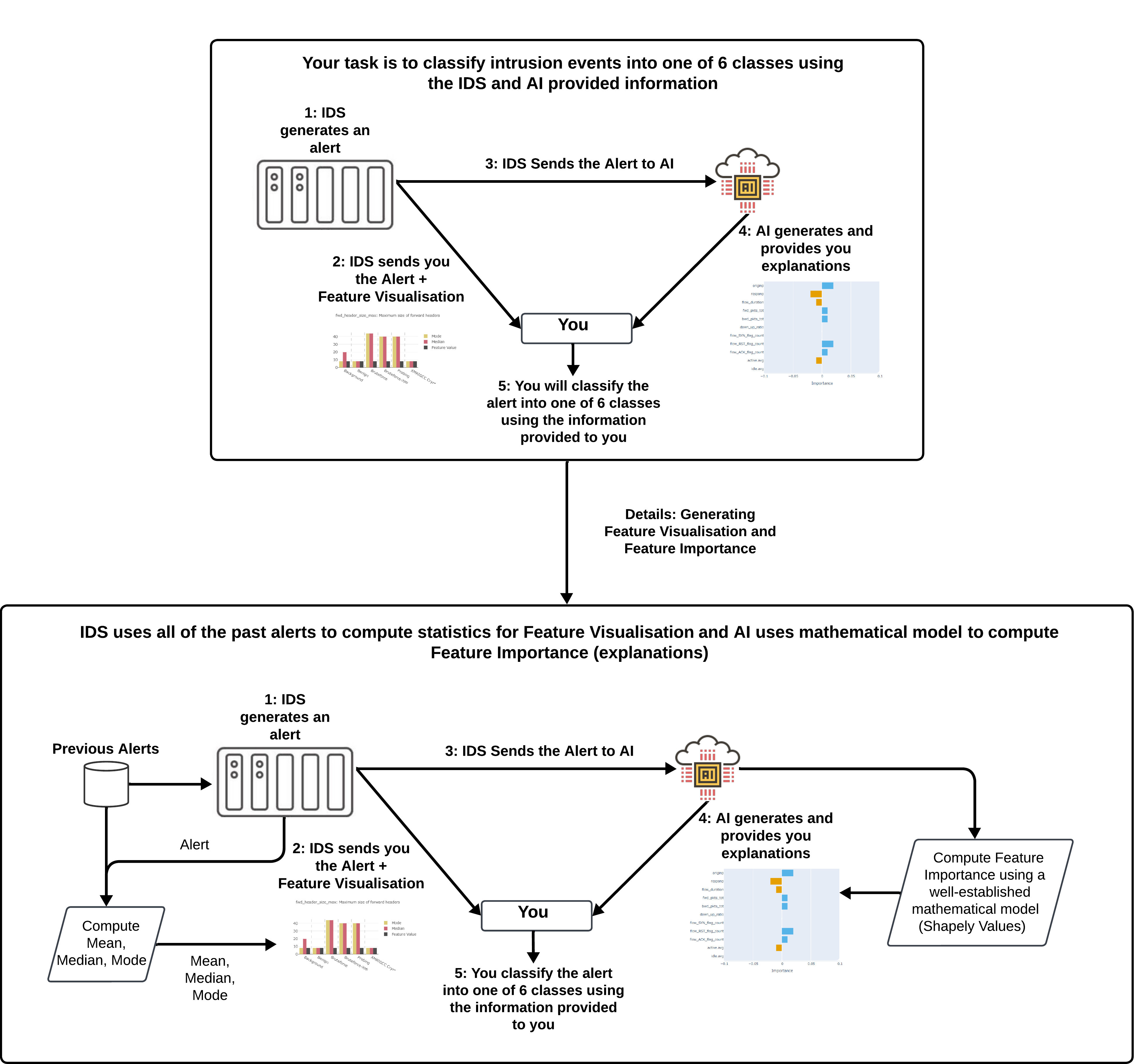}
    \caption{High-level overview of participant task.}
    \label{fig:task}
\end{figure*}

In this study, participants classified network intrusion alerts into one of six potential classes using feature statistic visualisations and AI-generated explanations (Figure~\ref{fig:task}). 

We provide an overview of participants' tasks and describe the dashboard elements they interact with during decision-making. The task flow involves the following key steps: 

\begin{enumerate} 
	\item \textbf{Alert Generation:} The dashboard presents four alerts per condition. 
	\item \textbf{Feature Visualisation:} The dashboard visualises features, including statistical summaries (mean, median, and mode). 

    \item \textbf{AI Explanations:} \systemname~ uses the existing library to compute feature importance. The dashboard provides participants with explanations of feature importance values for each class, indicating whether certain features positively or negatively support the alert being classified into each class. 

    \item \textbf{Classification:} Using the information provided (both feature visualisations and AI explanations), participants classify the alert into one of six classes: \textit{Background}, \textit{Benign}, \textit{BruteForce}, \textit{BruteForce-XML}, \textit{Probing}, and \textit{XMRIGCC CryptoMiner}. 

\end{enumerate} 

\FloatBarrier

\subsection{User Study Dashboard Design}
\label{sec:user_study_dashboard}

The dashboard provides participants with multiple tools to assist in their decision-making process.

\subsubsection{Alerts} 
\begin{figure*}
    \centering
    \includegraphics[width=1.0\linewidth]{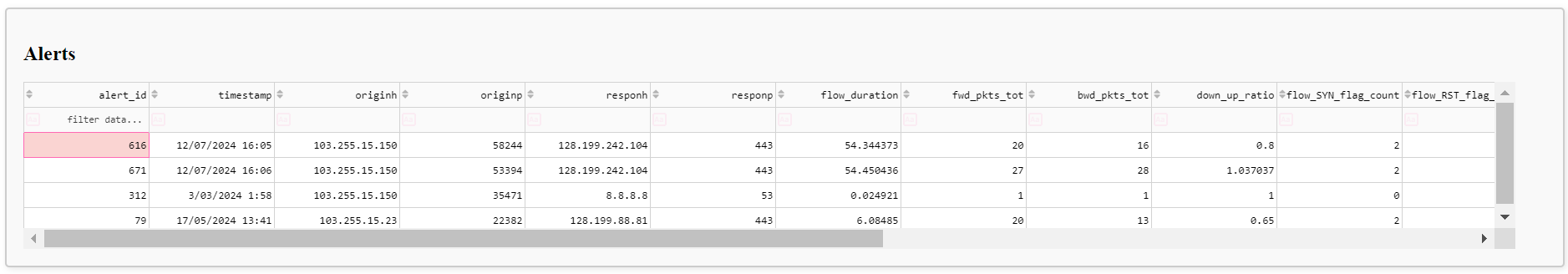}
    \caption{The alerts are displayed in a table.}
    \label{fig:panel1-alert-panel}
\end{figure*}

Participants begin by reviewing the raw alert data, which presents key features such as \textit{flow duration}, \textit{packet counts}, and \textit{flag counts} (Figure~\ref{fig:panel1-alert-panel}). 

\subsubsection{Feature Visualisation} 
\begin{figure*}
    \centering
    \includegraphics[width=1.0\linewidth]{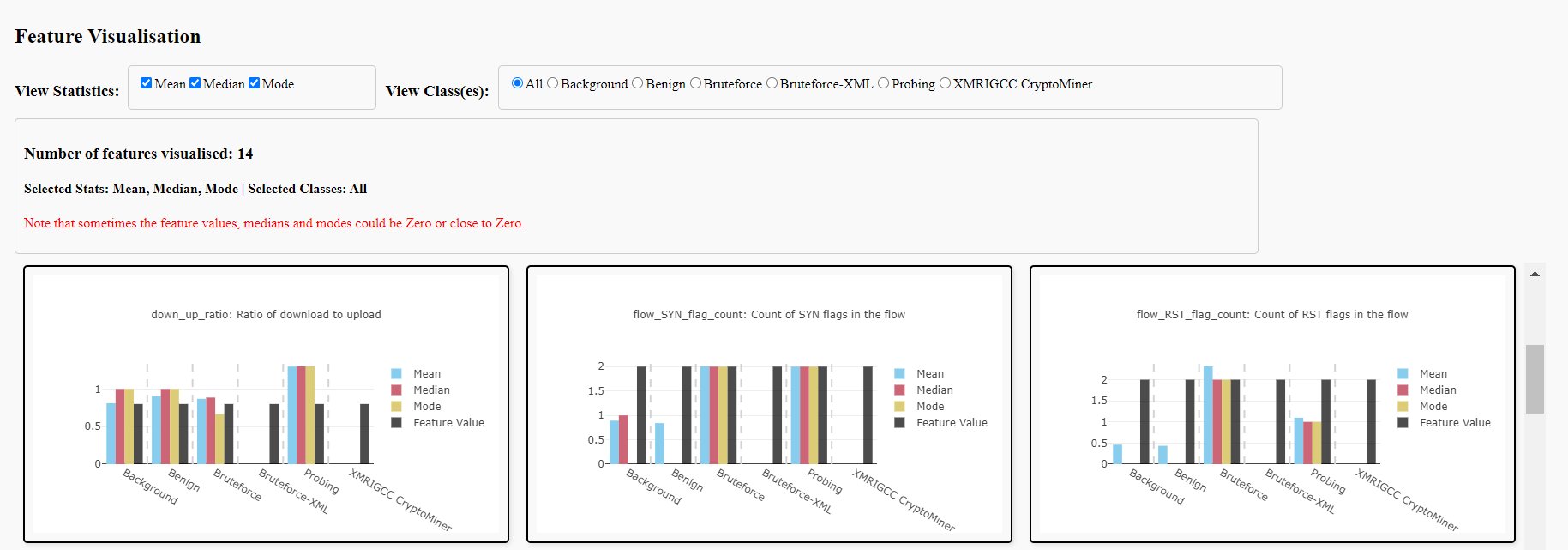}
    \caption{Example of feature visualisation showing key statistics (mean, median, and mode) for an alert.}
    \label{fig:panel2-fv-panel}
\end{figure*}

\begin{figure*}
    \centering
    \includegraphics[width=1.0\linewidth]{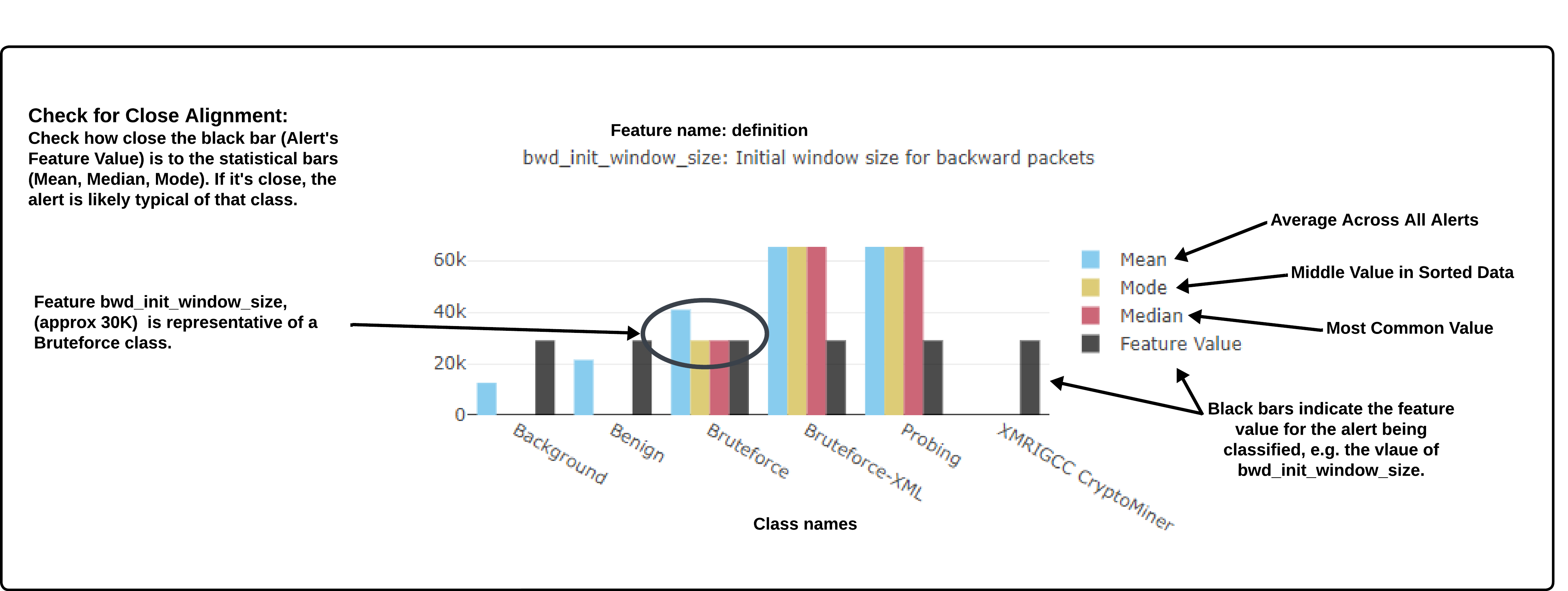}
    \caption{Explanation of feature visualisation.}
    \label{fig:fv-explanation}
\end{figure*}

The visualisation compares the alert's feature values with statistical summaries computed from past alerts (Figure~\ref{fig:panel2-fv-panel}). See Figure~\ref{fig:fv-explanation} to see how participants were told they could use it.

\subsubsection{Interaction Panel} 
\begin{figure*}
    \centering
    \includegraphics[width=1.0\linewidth]{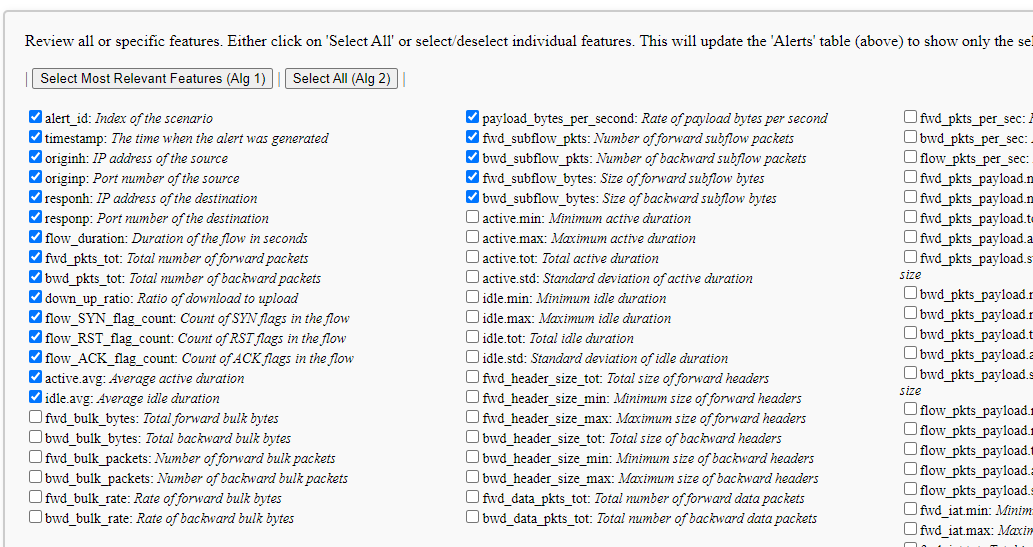}
    \caption{Iterative Panel: Allows participants to filter features. This panel is present in C3: Iterative assistance condition only.}
    \label{fig:panel3-interaction-panel}
\end{figure*}

This panel provides iterative assistance (only in Iterative condition) by allowing participants to selectively filter and focus on all, \systemname~selected, or participant-selected features, which are then visualised and explained in a more tailored manner.  (Figure~\ref{fig:panel3-interaction-panel}).

\subsubsection{AI-Generated Explanations} 
\begin{figure*}
    \centering
    \includegraphics[width=1.0\linewidth]{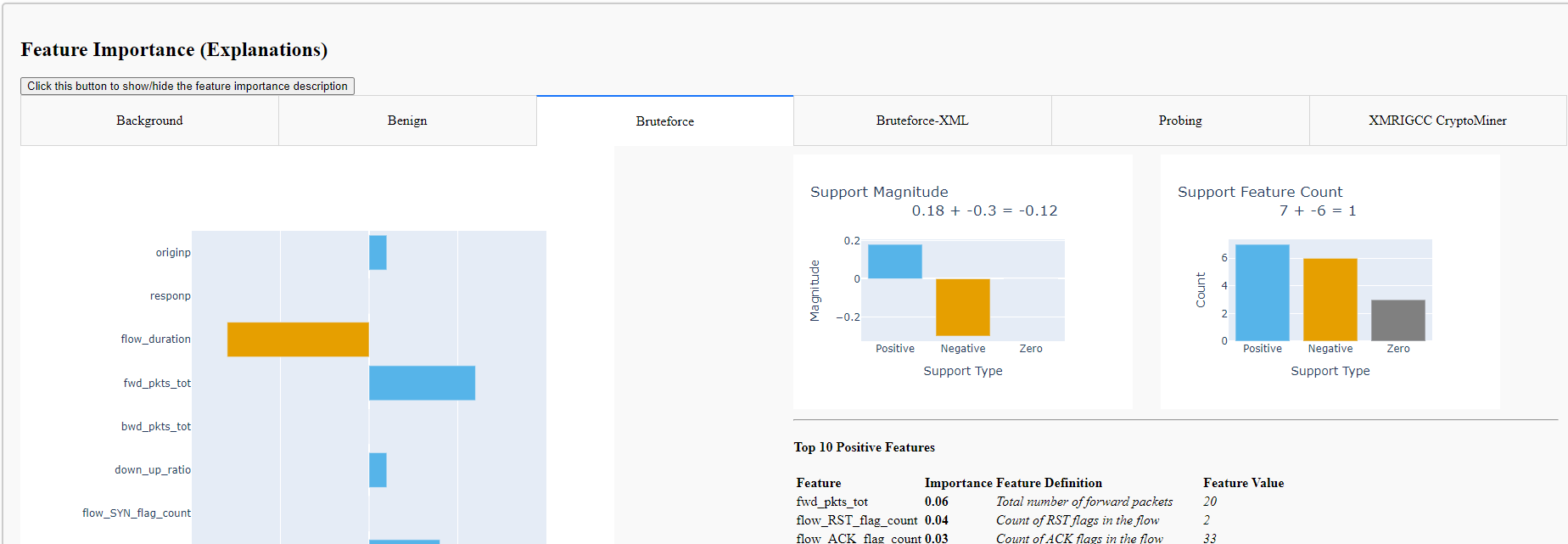}
    \caption{AI-generated explanations displaying feature importance values for different classes.}
    \label{fig:panel4-xai-panel}
\end{figure*}

Alongside the feature visualisations, the AI provides explanations in the form of feature importance values, computed using Shapley values. These explanations show each feature's positive or negative contribution towards the alert being classified into each class (Figure~\ref{fig:panel4-xai-panel}).

\begin{figure*}
    \centering
    \includegraphics[width=1.0\linewidth]{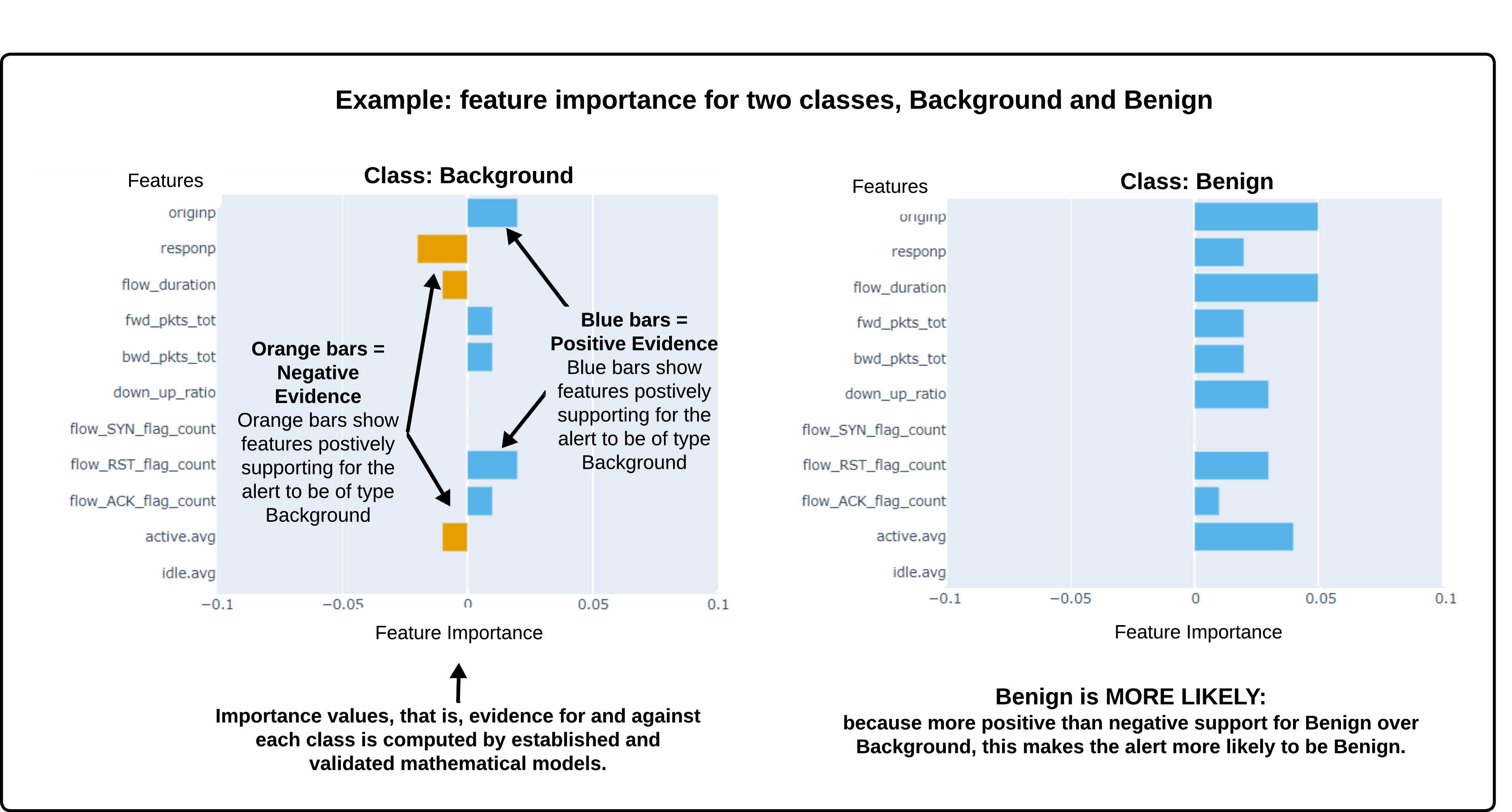}
    \caption{Discussion of how to interpret explainability information.}
    \label{fig:xai-explanation}
\end{figure*}

\subsubsection{Final Classification Decision}

\begin{figure*}
    \centering
    \includegraphics[width=1.0\linewidth]{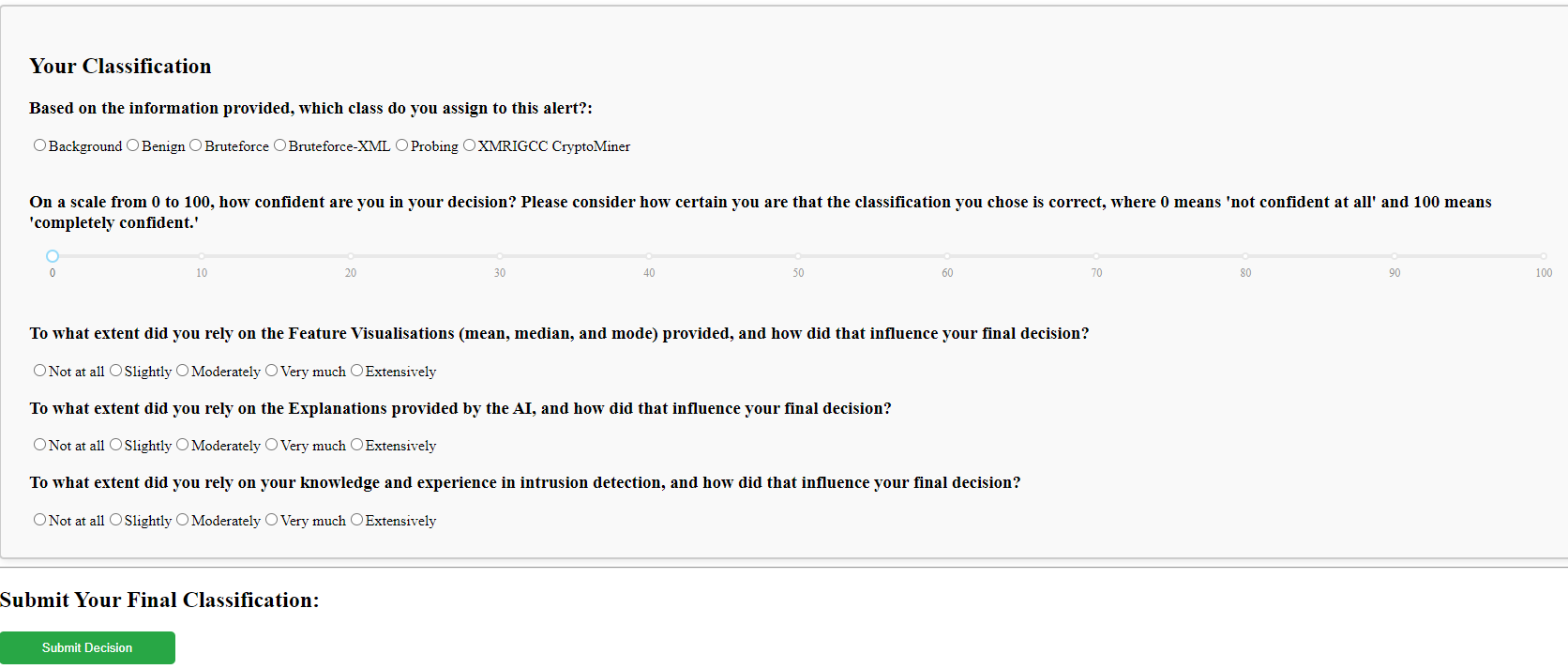}
    \caption{Final classification decision panel where participants submit their classification and report confidence.}
    \label{fig:panel5-task-panel}
\end{figure*}

After reviewing the feature visualisations and AI explanations, participants are asked to classify the alert into one of six potential classes. They will also report their confidence level in the decision and indicate to what extent they relied on the visualisations, explanations, and prior knowledge (Figure~\ref{fig:panel5-task-panel}). 

\subsection{Results}

\begin{figure}
    \centering
    \includegraphics[width=0.8\linewidth]{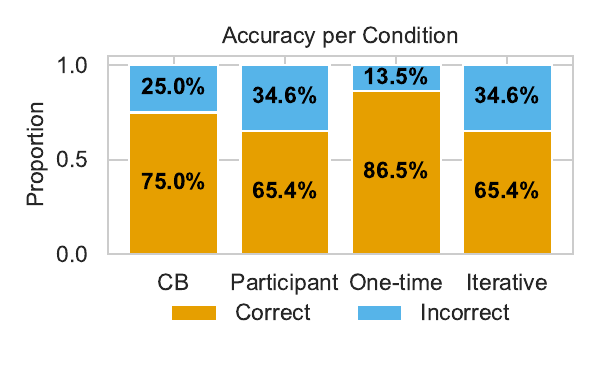}
    \caption{User accuracies.}
    \label{fig:stacked_accuracy_bars_colorblind}
\end{figure}

\begin{figure}
    \centering
    \includegraphics[width=0.8\linewidth]{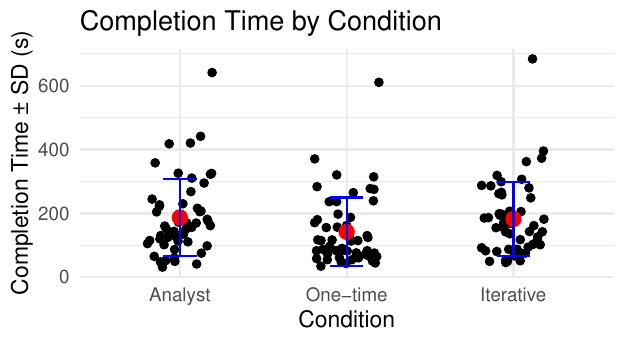}
    \caption{Completion times.}
    \label{fig:completion_time}
\end{figure}

\begin{figure}
    \centering
    \includegraphics[width=0.8\linewidth]{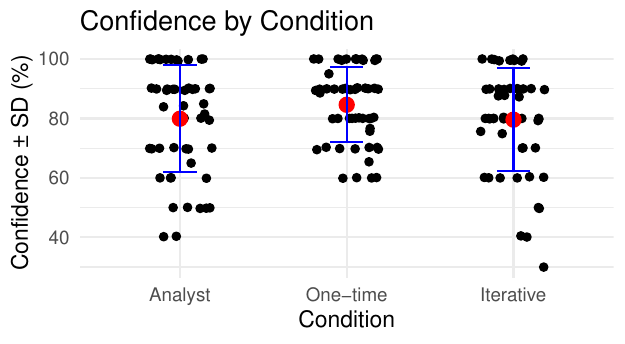}
    \caption{User confidence.}
    \label{fig:confidence}
\end{figure}

\begin{figure}
    \centering
    \includegraphics[width=0.80\linewidth]{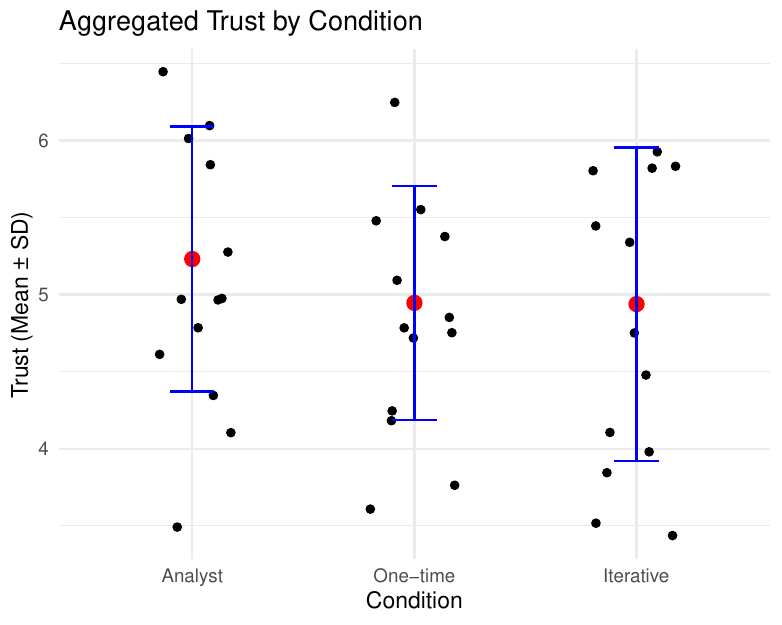}
    \caption{Aggregated trust.}
    \label{fig:trust}
\end{figure}

\begin{figure}
    \centering
    \includegraphics[width=0.8\linewidth]{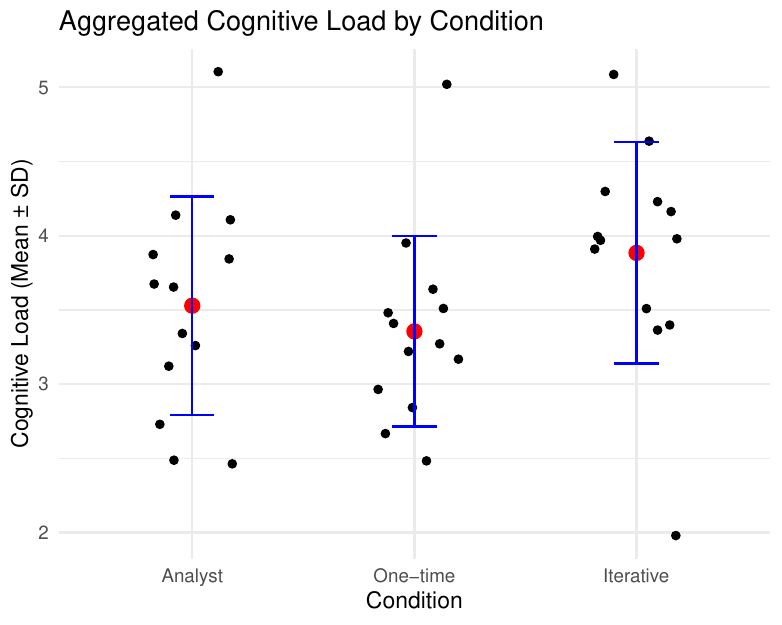}
    \caption{Aggregated cognitive load.}
    \label{fig:cognitive_load}
\end{figure}


Participants in the study completed two questionnaires: Trust in AI Explanations and Perceived Cognitive Load. 

\subsubsection{Trust in AI Explanations} Participants were asked to express their trust in the AI and its provided explanations. We explicitly instructed them that they were rating the AI and the explanations that it generates relative to the task and not their trust in AI in general. They rated their trust on a 7-point Likert scale, where 1 corresponds to "Not at all" and 7 corresponds to "Extremely." The items included in the Trust in AI questionnaire are as follows:
\begin{itemize}
    \item The AI is deceptive (could mislead, e.g., by providing false or misleading information).
    \item The AI behaves in an underhanded manner (operates in a way that may be dishonest).
    \item I am suspicious of the AI's intent, actions, or outputs.
    \item I am wary of the AI (I am cautious or hesitant when using the AI).
    \item The AI's actions will have a harmful or injurious outcome.
    \item I am confident in the AI.
    \item The AI is dependable.
    \item The AI is reliable.
    \item I can trust the AI.
    \item I am familiar with the AI.
\end{itemize}

\subsubsection{Perceived Cognitive Load}

Participants were also asked to express their perceived cognitive load while performing the task of classifying intrusion alerts. The cognitive load was rated on a 7-point scale, with 1 being "Very low" and 7 being "Very high." The items included in the Perceived Cognitive Load questionnaire are as follows:

\begin{itemize}
    \item For this task, many things needed to be kept in mind simultaneously.
    \item This task was very complex.
    \item I made an effort, not only to understand several details, but to understand the overall context.
    \item My point while dealing with the task was to understand everything correctly.
    \item The learning task consisted of elements supporting my comprehension of the task.
    \item During this task, it was exhausting to find the important information.
    \item The design of this task was very inconvenient for learning.
    \item During this task, it was difficult to recognise and link the crucial information.
\end{itemize}


\subsection{Alerts used in the User Study}
\small

\begin{table*}[ht]
\centering
\begin{adjustbox}{max width=\textwidth}
\begin{tabular}{ccccccccccc}
\toprule
\textbf{dataset} & \textbf{alert\_id} & \textbf{Bruteforce-XML} & \textbf{Bruteforce} & \textbf{Background} & \textbf{Benign} & \textbf{Probing} & \textbf{CryptoMiner} & \textbf{predicted\_class} & \textbf{ground\_truth} & \textbf{classification} \\
\midrule
1 & 476 & 0 & 0 & 0 & 0.04 & 0.96 & 0 & Probing & Probing & TP \\
1 & 625 & 0.07 & 0.02 & 0.29 & 0.42 & 0.2 & 0 & Benign & Probing & FN \\
1 & 279 & 0 & 0 & 0 & 1 & 0 & 0 & Benign & Benign & TN \\
1 & 640 & 0 & 0.39 & 0.38 & 0.22 & 0.01 & 0 & Bruteforce & Background & FP \\
2 & 284 & 0 & 0 & 0.01 & 0.99 & 0 & 0 & Benign & Benign & TN \\
2 & 91 & 0 & 0 & 0 & 0.02 & 0.98 & 0 & Probing & Probing & TP \\
2 & 674 & 0.01 & 0.4 & 0.33 & 0.24 & 0.02 & 0 & Bruteforce & Background & FP \\
2 & 664 & 0.01 & 0.05 & 0.16 & 0.41 & 0.37 & 0 & Benign & Probing & FN \\
3 & 79 & 0 & 1 & 0 & 0 & 0 & 0 & Bruteforce & Bruteforce & TP \\
3 & 616 & 0.02 & 0.03 & 0.26 & 0.13 & 0.56 & 0 & Probing & Benign & FP \\
3 & 312 & 0 & 0 & 0 & 1 & 0 & 0 & Benign & Benign & TN \\
3 & 671 & 0.02 & 0.04 & 0.47 & 0.17 & 0.3 & 0 & Background & Probing & FN \\
\bottomrule
\end{tabular}
\end{adjustbox}
\caption{Datasets used for user studies. It shows the distribution of alerts across the three datasets, each containing 4 alerts. For each alert, we show the dataset and alert ID, predicted probabilities of the random forest classifier (also used by the simulated analysts), predicted class and the ground truth from the dataset, and the classification outcomes. Note that we counterbalanced the order of datasets 1 and 2 across conditions C1 and C2 so that the noted improvements by \systemname~ were not due to specific instances in dataset 1 or 2. Dataset 3 was always used with C3.   }
\end{table*}

\normalsize


\subsection{Training Simulated Analysts, \systemname, and Testing the Dyad}

In this section, we describe the process of training simulated analysts, training \systemname, and testing the collaboration between the human analyst and \systemname~ in a SOC environment. The process is divided into three phases: simulated analyst training, \systemname~ training, and testing.

\subsection{Simulated Analyst Training}

The first phase involves training simulated analysts in a simulated SOC environment. Analysts interact with alerts by requesting features and additional contextual information to classify the nature of the alert. Each alert is investigated multiple times to ensure that the simulated analysts learn to classify the historical data correctly.

\begin{figure*}[h!]
    \centering
    \includegraphics[width=\textwidth]{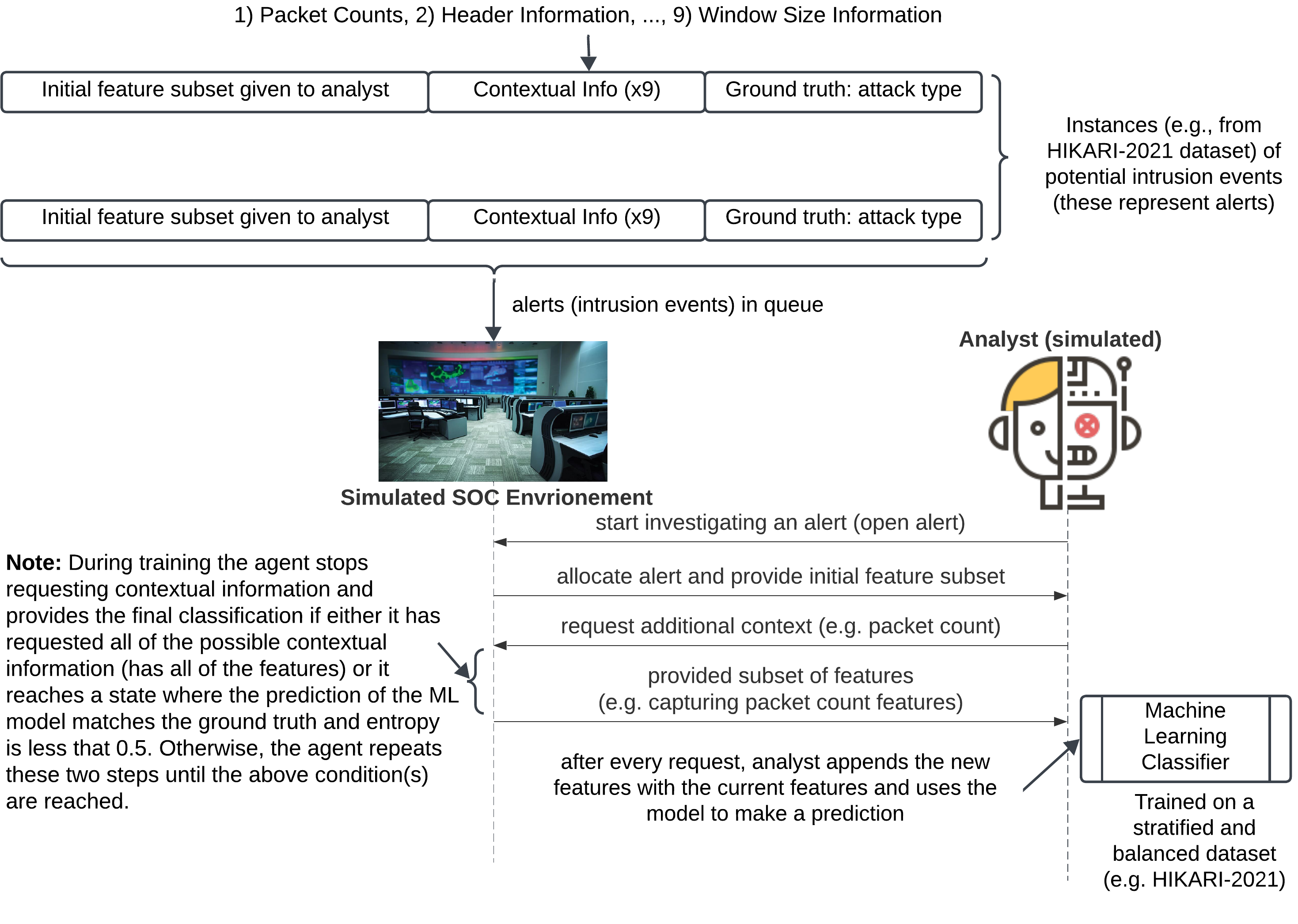}
    \caption{Simulated Analyst Training Process}
    \label{fig:analyst_training}
\end{figure*}

Figure \ref{fig:analyst_training} shows the key steps in this training process:

\begin{enumerate}
    \item The analyst is allocated an alert and provided with an initial subset of features.
    \item Based on this initial information, the analyst requests additional context (e.g., packet count) as needed.
    \item The requested subset of features is returned to the analyst.
    \item The analyst appends the additional features to the initial subset and uses a machine learning model (e.g., XGBoost classifier) to predict the alert type.
\end{enumerate}

This iterative process allows the simulated analysts to learn how to classify alerts efficiently by requesting the most relevant context features. The ground truth for the alerts (the actual attack type recorded in the dataset) is used to verify the accuracy of the analysts' predictions.

\subsection{Training \systemname{} (example with AIRL)}

The next phase involves training \systemname, which uses Adversarial Inverse Reinforcement Learning (AIRL) to mimic the decision-making process of the human analysts. ContextBuddy learns from the actions of the simulated analysts and the context features they request during their investigations.

\begin{figure*}[h!]
    \centering
    \includegraphics[width=\linewidth]{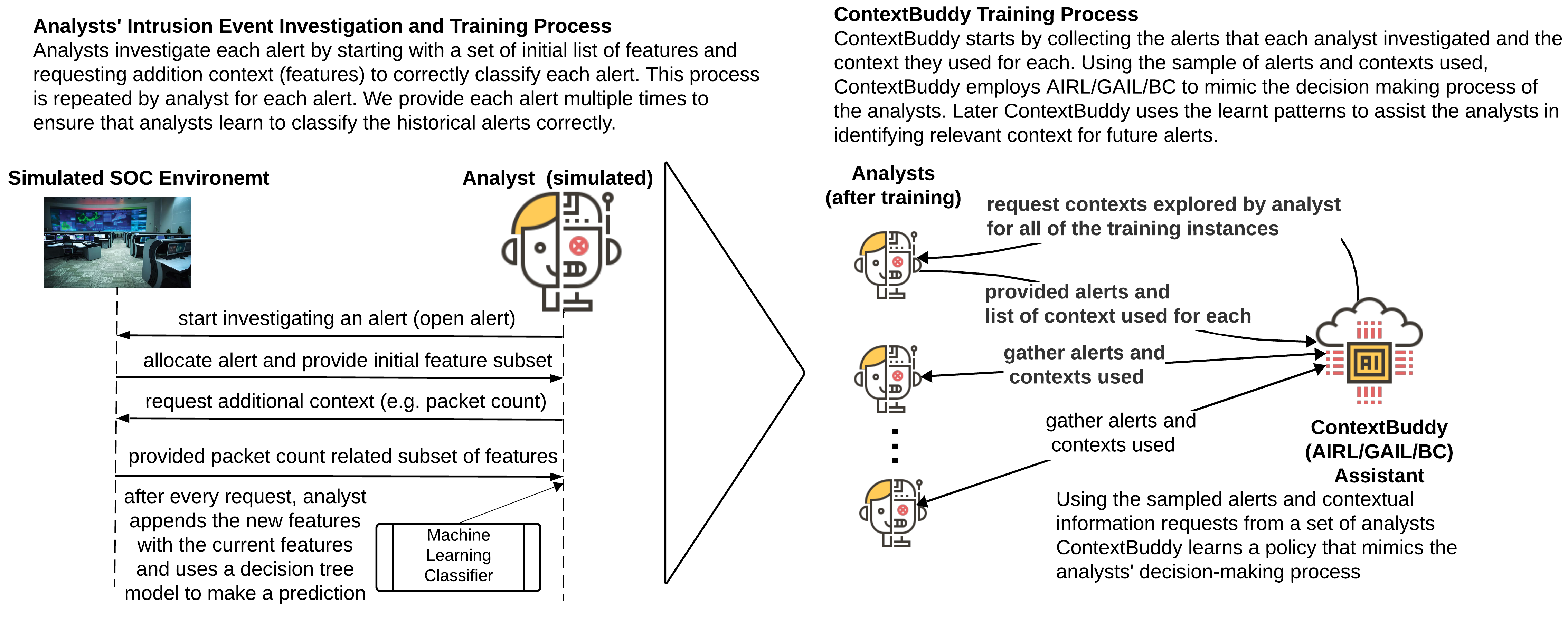}
    \caption{ContextBuddy (AIRL Assistant) Training Process}
    \label{fig:contextbuddy_training}
\end{figure*}

Figure \ref{fig:contextbuddy_training} illustrates the training process:

\begin{enumerate}
    \item The system collects the alerts and context features that the analysts explored during the training phase.
    \item These data are used to train \systemname, which employs AIRL to learn a policy that mimics the analysts' decision-making processes.
    \item \systemname~ generalises from this training data and develops the ability to identify the most relevant context features for future alerts.
\end{enumerate}

After training, \systemname~ can assist analysts by suggesting relevant context features for new alerts, effectively streamlining the classification process.

\subsection{Testing the Simulated Analysts and \systemname~ Dyad}

Once the simulated analysts and \systemname~ are trained, the next step is testing their collaboration. In this phase, the simulated analyst and \systemname~ classify new alerts in a SOC environment. The SOC provides an initial subset of features to the analyst, who can request additional context based on their investigation.

\begin{figure*}[h!]
    \centering
    \includegraphics[width=\textwidth]{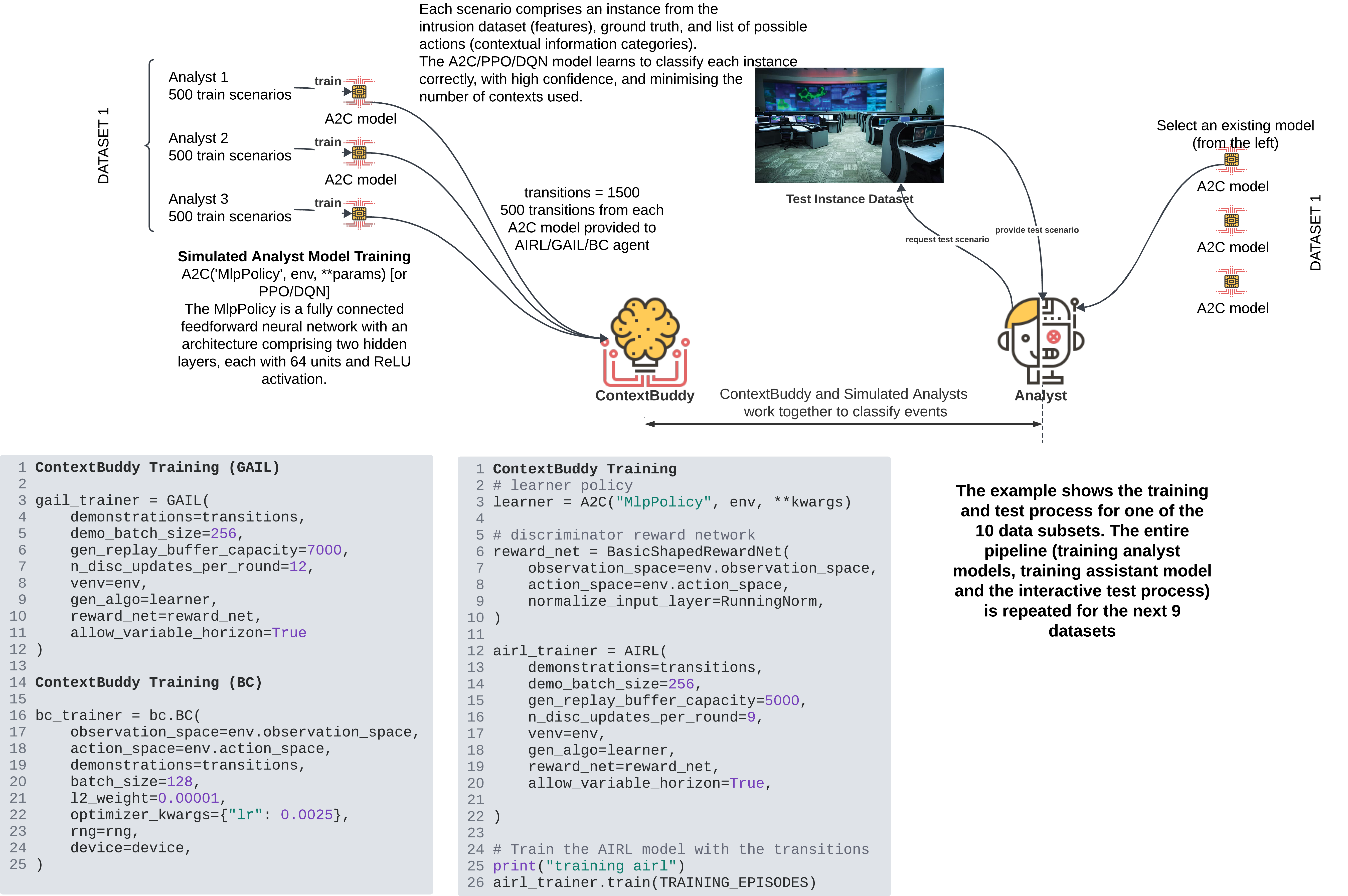}
    \caption{Testing the Simulated Analyst-\systemname~ Dyad in simulation-based experiments.}
    \label{fig:human_ai_dyad}
\end{figure*}

Figure \ref{fig:human_ai_dyad} shows the testing workflow:

\begin{enumerate}
    \item The SOC agent allocates an alert and provides the simulated analyst with an initial subset of features.
    \item The analyst can request further contextual information as needed, while \systemname~ assists by suggesting the most relevant context based on its learned policy.
    \item The analyst makes a final decision about the alert classification using a machine learning classifier.
\end{enumerate}

This collaboration ensures that the analyst and \systemname~ contribute to the alert classification process, leveraging each party's strengths. \systemname~ assists by reducing the analyst's cognitive load and offering contextual suggestions that will likely improve classification accuracy.

\subsection{Additional Results on Teamwork Between Simulated Analysts and \systemname}
The following section provides additional results for the simulation-based study that we could not include in the main paper due to space limitations.

\subsubsection{Breakdown of Simulated Analysts' Performance by Individual Analysts}
The following section provides a breakdown of analysts' performance by individual analyst instances. Recall that the results reported in the main paper are aggregated results.

\begin{figure}
    \centering
    \includegraphics[width=\linewidth]{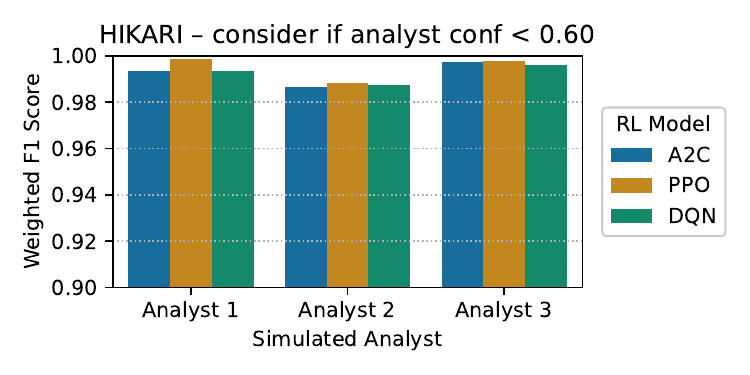}
    \caption{Breakdown of performance by individual analysts when threshold was 0.6 (Hikari)}    \label{fig:f1_by_analyst_hikari_consider_if_analyst_conf_lt_0.60}
\end{figure}

\begin{figure}
    \centering
    \includegraphics[width=\linewidth]{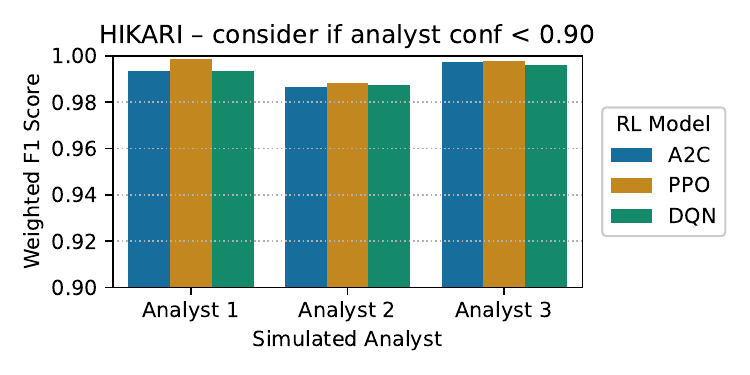}
    \caption{Breakdown of performance by individual analysts when threshold was 0.9 (Hikari)}    \label{fig:f1_by_analyst_hikari_consider_if_analyst_conf_lt_0.90}
\end{figure}

\begin{figure}
    \centering
    \includegraphics[width=\linewidth]{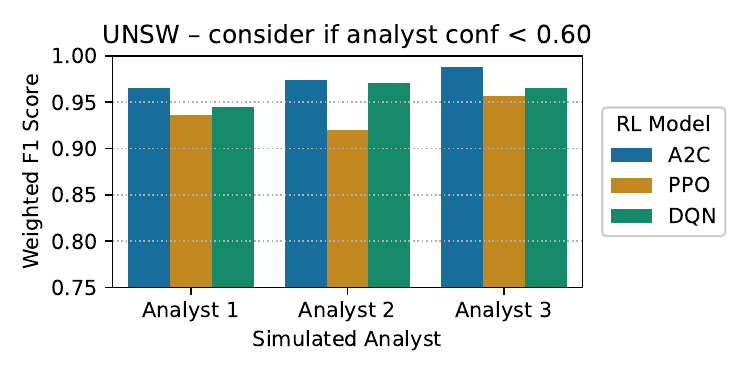}
    \caption{Breakdown of performance by individual analysts when threshold was 0.6 (UNSW)}    \label{fig:f1_by_analyst_unsw_consider_if_analyst_conf_lt_0.60}
\end{figure}

\begin{figure}
    \centering
    \includegraphics[width=\linewidth]{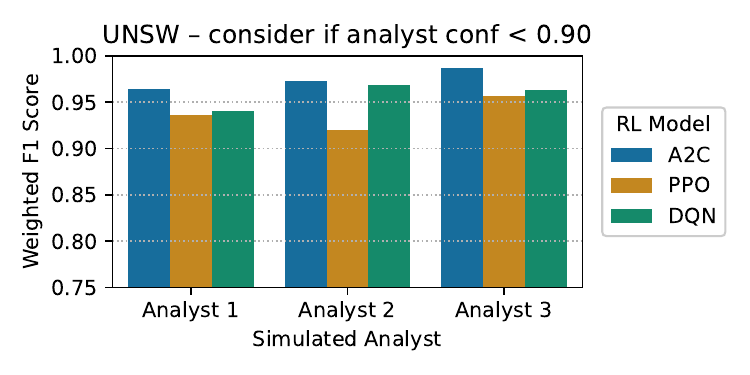}
    \caption{Breakdown of performance by individual analysts when threshold was 0.9 (UNSW)}    \label{fig:f1_by_analyst_unsw_consider_if_analyst_conf_lt_0.90}
\end{figure}

\subsubsection{Analysts' Confidence When Working Alone vs With \systemname}
The following plots show the effect on prediction confidence when simulated analysts were assisted by \systemname{}

\begin{figure}
    \centering
    \includegraphics[width=\linewidth]{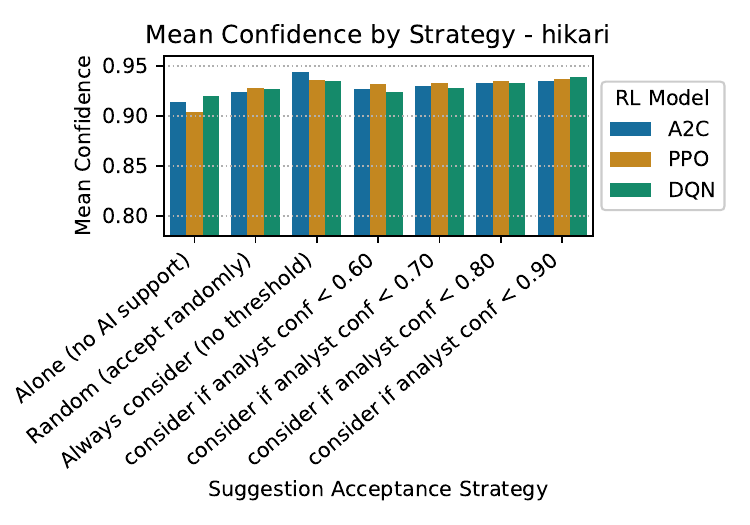}
    \caption{The analysts' prediction confidence (Hikari)}
    \label{fig:mean_conf_teaming_barplot_hikari}
\end{figure}

\begin{figure}
    \centering
    \includegraphics[width=\linewidth]{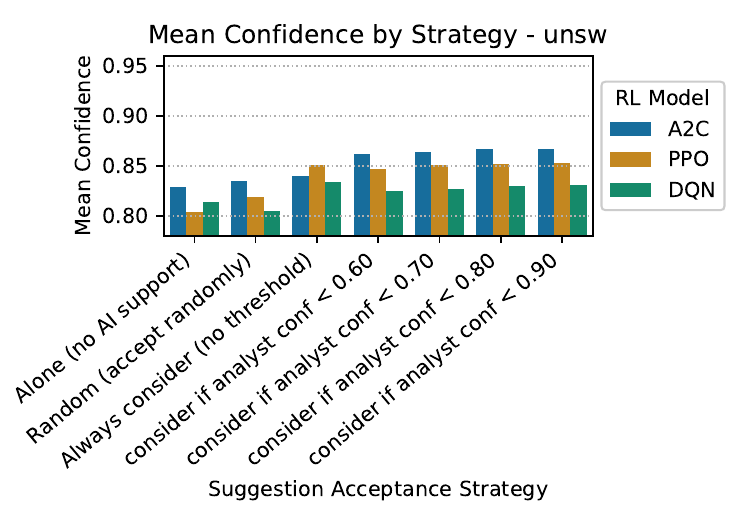}
    \caption{The analysts' prediction confidence (UNSW)}
    \label{fig:mean_conf_teaming_barplot_unsw}
\end{figure}

\end{document}